\documentclass[letterpaper,twocolumn,final]{IEEEtran}

\usepackage[twoside, left=1in, right=1in, top=1.0in, bottom=1.0in]{geometry}
\usepackage{amsmath}
\usepackage{amsthm}
\usepackage{amssymb} 
\usepackage{cleveref}
\usepackage{aliascnt}
\usepackage{graphicx}
\usepackage{pgfplots}
\usepackage{bbm}
\renewenvironment{proof}{\begin{IEEEproof}}{\end{IEEEproof}}

\newtheorem{theorem}{Theorem}
\newtheorem{definition}{Definition}
\newtheorem{remark}{Remark}
\newtheorem{example}{Example}

\newaliascnt{axiom}{theorem}

\aliascntresetthe{axiom}

\newaliascnt{lemma}{theorem}
\newtheorem{lemma}[lemma]{Lemma}
\aliascntresetthe{lemma}

\newaliascnt{prop}{theorem}

\aliascntresetthe{prop}

\newaliascnt{corollary}{theorem}
\newtheorem{corollary}[corollary]{Corollary}
\aliascntresetthe{corollary}

\newaliascnt{conjecture}{theorem}

\aliascntresetthe{conjecture}

\newaliascnt{observation}{theorem}

\aliascntresetthe{observation}

\newaliascnt{algo}{theorem}

\aliascntresetthe{algo}

\newaliascnt{notation}{definition}

\aliascntresetthe{notation}

\crefname{equation}{}{}
\Crefname{equation}{}{}
\crefname{thm}{theorem}{theorems}
\Crefname{thm}{Theorem}{Theorems}
\crefname{app}{appendix}{appendices}
\Crefname{app}{Appendix}{Appendices}
\crefname{prop}{proposition}{propositions}
\Crefname{prop}{Proposition}{Propositions}
\crefname{figure}{fig.}{figures}
\Crefname{figure}{Fig.}{Figures}
\crefname{defn}{definition}{definitions}
\Crefname{defn}{Definition}{Definitions}
\crefname{fact}{fact}{facts}
\Crefname{fact}{Fact}{Facts}
\crefname{appendix}{appendix}{appendices}
\Crefname{appendix}{Appendix}{Appendices}
\crefname{algo}{algorithm}{algorithms}
\Crefname{algo}{Algorithm}{Algorithms}
\crefname{algorithm}{algorithm}{algorithms}
\Crefname{algorithm}{Algorithm}{Algorithms}
\crefname{conjecture}{conjecture}{conjectures}
\Crefname{conjecture}{Conjecture}{Conjectures}
\crefname{obs}{observation}{observations}
\Crefname{obs}{Observation}{Observations}

\newcommand{\addplot}[1]{\includegraphics[width=\linewidth,ext=.pdf,type=pdf,read=*]{#1}}

\usepackage{xargs}

\newcommand{\remove}[1]{}

\newcommand{\olfrac}[2]{{#1}/{#2}}
\newcommand{\oltfrac}[2]{{\left(#1\right)}/{#2}}
\newcommand{\olbfrac}[2]{{#1}/{\left(#2\right)}}

\newenvironment{olequation*}{\begin{math}}{\end{math}}

\newcommand{\ie}{i.e.}

\newcommand{\eg}{e.g.}

\newcommand{\cf}{cf.}

\newcommand{\whp}{{w.h.p.}}

\newcommand{\wrt}{{w.r.t.{}}}

\DeclareMathOperator*{\argmin}{arg\,min}

\DeclareMathOperator{\lesim}{\,\genfrac{}{}{0pt}{}{\raisebox{-5pt}{$<$}}{\raisebox{5pt}{\resizebox{7pt}{3pt}{$\propto$}}}\,}

\newcommand{\cardinality}[1]{\left|#1\right|}
\newcommand{\abs}[1]{\left|#1\right|}

\newcommand{\ntoinf}[1][n]{{#1}\rightarrow\infty}

\newcommand{\defeq}{\triangleq}

\newcommand{\setst}{:}
\newcommand{\SetDef}[2]{\left\{#1 \setst #2 \right\}}

\newcommand{\preals}{\mathbb{R}^+}

\newcommand{\pints}{\mathbb{Z^+}}

\newcommand{\eps}{\varepsilon}

\newcommand{\ceil}[1]{\left\lceil{#1}\right\rceil}
\newcommand{\floor}[1]{\left\lfloor{#1}\right\rfloor}

\newcommand{\onenorm}[1]{\left\|#1\right\|_1}

\newcommand{\infnorm}[1]{\left\|#1\right\|_{\infty}}

\newcommand{\BigO}[1]{O\left({#1}\right)}
\newcommand{\SmallO}[1]{o\left({#1}\right)}
\newcommand{\BigOmega}[1]{\Omega\left({#1}\right)}
\newcommand{\BigTheta}[1]{\Theta\left({#1}\right)}

\newcommand{\nchoosek}[2]{{{#1}\choose{#2}}}

\newcommand{\E}[1]{\mathbb{E}\left[{#1}\right]}
\newcommand{\Ep}[2]{\mathbb{E}_{#1}\left[{#2}\right]}

\newcommand{\Var}[1]{\mathrm{Var}\left[{#1}\right]}
\newcommand{\Std}[1]{\mathrm{Std}\left[{#1}\right]}

\newcommand{\PDUniform}{\textsf{Unif}}

\newcommand{\PUniformSet}[1]{\PDUniform\left({#1}\right)}

\newcommand{\Set}[1]{\left\{{#1}\right\}}

\newcommand{\Entropy}[1]{H\left(#1\right)}
\newcommand{\CondEntropy}[2]{H\left(#1|#2\right)}
\newcommand{\Hb}[1]{H_b\left(#1\right)}

\newcommand{\Prob}[1]{\mathbb{P}\left[{#1}\right]}

\newcommandx{\OSn}[3][1=X,3=n,usedefault]{{#1}_{{#2}:{#3}}}

\newcommand{\nn}{\nonumber}

\DeclareMathAlphabet{\mathbsf}{OT1}{cmss}{bx}{n}
\DeclareMathAlphabet{\mathssf}{OT1}{cmss}{m}{sl}

\DeclareSymbolFont{bsfletters}{OT1}{cmss}{bx}{n}  
\DeclareSymbolFont{ssfletters}{OT1}{cmss}{m}{n}
\DeclareMathSymbol{\bsfGamma}{0}{bsfletters}{'000}
\DeclareMathSymbol{\ssfGamma}{0}{ssfletters}{'000}
\DeclareMathSymbol{\bsfDelta}{0}{bsfletters}{'001}
\DeclareMathSymbol{\ssfDelta}{0}{ssfletters}{'001}
\DeclareMathSymbol{\bsfTheta}{0}{bsfletters}{'002}
\DeclareMathSymbol{\ssfTheta}{0}{ssfletters}{'002}
\DeclareMathSymbol{\bsfLambda}{0}{bsfletters}{'003}
\DeclareMathSymbol{\ssfLambda}{0}{ssfletters}{'003}
\DeclareMathSymbol{\bsfXi}{0}{bsfletters}{'004}
\DeclareMathSymbol{\ssfXi}{0}{ssfletters}{'004}
\DeclareMathSymbol{\bsfPi}{0}{bsfletters}{'005}
\DeclareMathSymbol{\ssfPi}{0}{ssfletters}{'005}
\DeclareMathSymbol{\bsfSigma}{0}{bsfletters}{'006}
\DeclareMathSymbol{\ssfSigma}{0}{ssfletters}{'006}
\DeclareMathSymbol{\bsfUpsilon}{0}{bsfletters}{'007}
\DeclareMathSymbol{\ssfUpsilon}{0}{ssfletters}{'007}
\DeclareMathSymbol{\bsfPhi}{0}{bsfletters}{'010}
\DeclareMathSymbol{\ssfPhi}{0}{ssfletters}{'010}
\DeclareMathSymbol{\bsfPsi}{0}{bsfletters}{'011}
\DeclareMathSymbol{\ssfPsi}{0}{ssfletters}{'011}
\DeclareMathSymbol{\bsfOmega}{0}{bsfletters}{'012}
\DeclareMathSymbol{\ssfOmega}{0}{ssfletters}{'012}

\newcommand{\cC}{\mathcal{C}}

\newcommand{\hD}{\hat{D}}

\newcommand{\brD}{\bar{D}}

\newcommand{\cG}{\mathcal{G}}

\newcommand{\cI}{\mathcal{I}}

\newcommand{\cM}{\mathcal{M}}

\newcommand{\hR}{\hat{R}}

\newcommand{\brR}{\bar{R}}

\newcommand{\cS}{\mathcal{S}}

\newcommand{\cV}{\mathcal{V}}

\newcommand{\bX}{\mathbf{X}}
\newcommand{\cX}{\mathcal{X}}

\newcommand{\ba}{\mathbf{a}}

\newcommand{\bd}{\mathbf{d}}

\newcommand{\hr}{\hat{r}}

\newcommand{\bx}{\mathbf{x}}

\newcommand{\by}{\mathbf{y}}

\newcommand{\hsigma}{\hat{\sigma}}

\newcommand{\sm}{\mathrm{s}}

\newcommand{\la}{\mathrm{l}}

\newcommand{\idperm}{{\mathrm{Id}}}

\newcommand{\Mallows}[2]{\cM\left({#1},{#2}\right)}
\newcommand{\MallowsId}[1]{\cM\left({#1}\right)}

\newcommand{\Ag}{A}

\newcommand{\RateA}{\brR}
\newcommand{\RateW}{\hR}
\newcommand{\RateG}{R}

\newcommand{\CodeG}[1][n]{{\cC}_{#1}}
\newcommand{\CodeA}[1][n]{\bar{\cC}_{#1}}
\newcommand{\CodeW}[1][n]{\hat{\cC}_{#1}}

\newcommand{\fsort}[1]{\mathrm{sort}\left({#1}\right)}

\newcommand{\invperm}[1][\sigma]{{#1}^{-1}}

\newcommand{\inv}[1]{\bx_{#1}}
\newcommand{\invx}[2]{\bx_{#1}\left({#2}\right)}
\newcommand{\einv}[1]{\tilde{\bx}_{#1}}
\newcommand{\einvx}[2]{\tilde{\bx}_{#1}\left({#2}\right)}

\newcommand{\ins}[1]{\ba_{#1}}
\newcommand{\insx}[2]{\ba_{#1}\left({#2}\right)}

\newcommand{\PermutationSet}[1][n]{\cS_{#1}}
\newcommand{\InvVectorSet}[1][n]{\cG_{#1}}
\newcommand{\PermutationSpaceText}{\cX}

\newcommand{\PermutationSpace}[2][\PermutationSet]{\PermutationSpaceText\left({#1},{#2}\right)}

\newcommand{\ktText}{\tau}
\newcommand{\ktdis}{Kendall tau distance}
\newcommand{\dtauText}{d_{\tau}}
\newcommand{\dtau}[2]{\dtauText\left({#1},{#2}\right)}

\newcommand{\LoneText}{\ell_1}
\newcommand{\LinfText}{\ell_{\infty}}

\newcommand{\spearman}{Spearman's footrule}
\newcommand{\invLoneName}{inversion-$\ell_1$ distance}
\newcommand{\invLoneText}{\bx,\ell_1}
\newcommand{\dinvLoneText}{d_{\invLoneText}}
\newcommand{\dinvLone}[2]{\dinvLoneText\left({#1},{#2}\right)}

\newcommand{\dLoneText}{d_{\ell_1}}
\newcommand{\dLone}[2]{\dLoneText\left({#1},{#2}\right)}

\newcommand{\dLinfText}{d_{\ell_\infty}}
\newcommand{\dLinf}[2]{\dLinfText\left({#1},{#2}\right)}

\newcommand{\fDtauA}[1]{\brD_{\dtauText}\left({#1}\right)}

\newcommand{\fDtauW}[1]{\hD_{\dtauText}\left({#1}\right)}

\newcommand{\fDLoneA}[1]{\brD_{\LoneText}\left({#1}\right)}

\newcommand{\fDLoneW}[1]{\hD_{\LoneText}\left({#1}\right)}

\newcommand{\fDLinfA}[1]{\brD_{\LinfText}\left({#1}\right)}

\newcommand{\fDLinfW}[1]{\hD_{\LinfText}\left({#1}\right)}

\newcommand{\DinvLoneW}{\hD_{\invLoneText}}
\newcommand{\fDinvLoneW}[1]{\hD_{\invLoneText}\left({#1}\right)}

\newcommand{\BallSize}[2][D]{N_{#2}\left({#1}\right)}
\newcommand{\Ball}[2][D]{B_{#2}\left({#1}\right)}

\newcommand{\LoneBallSize}[1][D]{\BallSize[#1]{\LoneText}}
\newcommand{\LoneBall}[1][D]{\Ball[#1]{\LoneText}}

\newcommand{\LinfBallSize}[1][D]{\BallSize[#1]{\LinfText}}

\newcommand{\invLoneBallSize}[1][D]{\BallSize[#1]{\invLoneText}}
\newcommand{\invLoneBall}[1][D]{\Ball[#1]{\invLoneText}}

\newcommand{\ktBallSize}[1][D]{\BallSize[#1]{\ktText}}
\newcommand{\ktBall}[1][D]{\Ball[#1]{\ktText}}

\begin{document}
\title{Compression in the Space of Permutations}

\author{
Da~Wang,
Arya~Mazumdar,~\IEEEmembership{Member,~IEEE,}
and 
Gregory~W.~Wornell,~\IEEEmembership{Fellow,~IEEE}%
\thanks{
This work was supported, in part, by AFOSR under Grant No.~FA9550-11-1-0183, and by NSF under
Grant No.~CCF-1017772 and Grant No.~CCF-1318093.}%
\thanks{
D. Wang was 
with the Department of Electrical Engineering and Computer Science, Massachusetts Institute of Technology, Cambridge, MA, 02139
and now with Two Sigma Investments, New York, NY, 10013 (Email: dawang@alum.mit.edu).
A. Mazumdar is with the Department of Electrical and Computer Engineering, University of Minnesota, MN, 55455 (Email: arya@umn.edu).
G. W. Wornell is
with the Department of Electrical Engineering and Computer Science, Massachusetts Institute of Technology, Cambridge, MA, 02139
(Email: gww@mit.edu).
}
\thanks{
The material in this paper was presented in part at 
International Symposium on Information Theory, Istanbul, Turkey, 2013 and 
International Symposium on Information Theory, Hawaii, HI, 2014. 
}
}




\maketitle

\begin{abstract}
    We investigate lossy compression (source coding) of data in the form of permutations. This problem has
    direct applications in the storage of ordinal data or rankings, and in the analysis of sorting algorithms.
    We analyze the rate-distortion characteristic for the permutation space under the uniform distribution, and
    the minimum achievable rate of compression that allows a bounded distortion after recovery.  Our analysis is
    with respect to different practical and useful distortion measures, including \ktdis{},
    Spearman's footrule, Chebyshev distance and \invLoneName. 
    We establish equivalence of source code designs under certain distortions and
    show simple explicit code designs that incur low encoding/decoding complexities and are asymptotically optimal.
    Finally, we show that for the Mallows model, a popular nonuniform ranking model on the permutation space,
    both the entropy and the maximum distortion at zero rate are much lower than the uniform counterparts, which
    motivates the future design of efficient compression schemes for this model.

\end{abstract}


\section{Introduction}
\IEEEPARstart{P}{ermutations} are fundamental mathematical objects and the topic of \emph{codes in permutations}
is a well-studied subject in coding theory. A variety of applications that correspond to different metric
functions on the symmetric group on $n$ elements $\PermutationSet$ have been investigated. For example, some
works focus on error-correcting codes in $\PermutationSet$ with Hamming
distance~\cite{blake_coding_1979,colbourn_permutation_2004}, and some others investigate the error correction
problem under metrics such as Chebyshev distance~\cite{klve_permutation_2010} and \ktdis~\cite{barg_codes_2010}.

While error correction problems in permutation spaces have been investigated before, the lossy compression
problem is largely left unattended. In \cite{barbay_compressed_2009,barbay_lrm-trees:_2012}, the authors
investigate the lossless compression of a group of permutations with certain properties, such as efficient rank
querying (given an element, get its rank in the permutation) and selection (given a rank, retrieve the
corresponding element). By contrast, in this paper we consider the lossy compression (source coding) of
permutations, which is motivated by the problems of storing ranking data, and lower bounding the complexity of
approximate sorting, which we now describe.

\textbf{Storing ranking data:}
In applications such as recommendation systems, users rank products and these rankings are analyzed to
provide new recommendations. To have personalized recommendation, it may be necessary to store the ranking data for
each user in the system, and hence the storage efficiency of ranking data is of interest. 
Because a ranking of $n$ items can be represented as a permutation of 1 to $n$, storing a ranking is equivalent to
storing a permutation. 
Furthermore, in many cases a rough knowledge of the ranking (\eg, finding one of the top five elements
instead of the top element) is sufficient. 
This poses the question of the number of bits needed for permutation storage when a certain amount of error can be
tolerated. 
In many current applications the cost of lossless storage is usually tolerable and hence lossy compression may
not be necessary. However lossy compression is a fundamental topic and it is of theoretical interest to
understand the trade-off involved.

\textbf{Lower bounding the complexity of approximate sorting:}
Given a group of elements of distinct values, comparison-based sorting can be viewed as the
process of searching for a true ranking by pairwise comparisons. Since each comparison in
sorting provides at most 1 bit of information, the log-size of the permutation set
$\PermutationSet$, $\log_2 (n!)$, provides a lower bound to the required number of comparisons.
Similarly, the lossy source coding of permutations provides a lower bound on the number of comparisons 
to the problem of comparison-based
approximate sorting, which can be seen as finding a true permutation up to a certain distortion. Again,
the log-size of the code indicates the amount of information (in  bits) needed to specify the true
permutation, which in turn provides a lower bound on the number of pairwise comparisons needed.

In one line of work, authors of 
\cite{giesen_approximate_2009} derived both lower and upper bounds for approximate sorting in some range of
allowed distortion with respect to the Spearman's footrule metric~\cite{diaconis_spearmans_1977} (see
\Cref{def:spearman_footrule} below). 
Another line of work concerns an important class of approximate sorting, the problem of \emph{partial sorting}, first proposed in \cite{chambers_algorithm_1971}
(\cf \cite[Chapter 8]{wang_computing_2014} for an exposition on the relationships between various sorting
problems).
Given a set of $n$ elements $\cV$ and a set of indices $\cI \subset \Set{1,
2, \ldots, n}$, a partial sorting algorithm aims to arrange the elements into a list $[v_1, v_2, \ldots, v_n]$
such that for any $i \in \cI$, all elements with indices $j < i$ are no greater than $v_i$, and all elements
with indices $j' > i$ are no smaller than $v_i$. A partial sorting algorithm essentially selects all elements
with ranks in the set $\cI$, and hence is also called \emph{multiple selection}.  
The information-theoretic lower bound for partial sorting algorithms have been proposed
in~\cite{fredman_how_1976}, and the authors of~\cite{kaligosi_towards_2005} propose a multiple selection
algorithms with expected number of comparisons within the information-theoretic lower-bound and an
asymptotically negligible additional term. 

Comparing with existing work (such as \cite{fredman_how_1976}),
our analysis framework via rate-distortion theory 
is more general
as we provide an information-theoretic lower bound on the query complexity for \emph{all} approximate sorting
algorithms that achieve a certain distortion, and the multiple selection algorithm proposed
in~\cite{kaligosi_towards_2005} turns out to be optimal for the general approximate sorting problem as well.
Therefore, our information-theoretic lower bound is tight.


\begin{remark}[Comparison-based sorting implies compression]
It is worth noting that every comparison-based sorting algorithm corresponds to a compression scheme of the
permutation space. In particular, the string of bits that represent comparison outcomes in any deterministic
(approximate) sorting algorithm corresponds to a (lossy) representation of the permutation. 

For a more in-depth discussion on the relationship between sorting and compression, see~\cite{barbay2013time} and references therein.
\end{remark}

Beyond the above applications, the rate-distortion theory in permutation spaces is of technical interest on its
own because the permutation space does not possess the product structure that a discrete memoryless source
induces.

With the above motivations, we consider the problem of lossy compression in permutation
spaces in this paper.
Following the classical rate-distortion formulation, we aim to determine, given a distortion measure $d(\cdot, \cdot)$, 
the minimum number of bits needed to describe a permutation with distortion at most $D$.

The analysis of the lossy compression problem depends on the source distribution and the distortion
measure. We are mainly concerned with the permutation spaces with a uniform distribution, and consider different distortion
measures based on four distances in the permutation spaces:
the \ktdis{}, Spearman's footrule, Chebyshev distance and \invLoneName.
As we shall see in \Cref{sec:rd_formulation}, each of these distortion measures (except \invLoneName%
\footnote{We are interested in \invLoneName\ due to its extremal property shown in Equation \eqref{eq:geq_rlships}, which is useful when we derive results for other permutation spaces. 
Further use of this metric in the context of smooth representation of permutations can be found in~\cite{mazumdar_smooth_2014}.
}%
) has its own operational meaning
that may be useful in different applications.

In addition to characterizing the trade-off between rate and distortion, we also show that under the uniform
distribution over the permutation space, there are close relationships between some of the distortion measures of
interest in this paper. We use these relations to establish the corresponding equivalence of source codes in permutation
spaces with different distortion measures. 
For each distortion measure, we provide simple and constructive achievability
schemes, leading to explicit code designs with low complexity.

Finally, we turn our attention to non-uniform distributions over the permutation space. In some applications,
we may have prior knowledge about the permutation data, which can be captured in a model of non-uniform distribution.
There are a variety of distributional models in different contexts, such as the Bradley-Terry model~\cite{bradley_rank_1952}, the
Luce-Plackett model~\cite{luce_individual_1959,plackett_analysis_1975}, and the Mallows
model~\cite{mallows_non-null_1957}. Among these, we choose the Mallows model due
to its richness and applicability in various ranking
applications~\cite{cheng_new_2009,klementiev_unsupervised_2008,lu_learning_2011}.
We analyze the lossless and lossy compression of the permutation space under the Mallows model and with the \ktdis{}
as the distortion measure, and characterize its entropy and end points of its rate-distortion function.

The rest of the paper is organized as follows. We first present the problem
formulation in \Cref{sec:rd_formulation}. We then analyze the geometry of the permutation spaces and show that there exist close relationships
between some distortion measures of interest in this paper in
\Cref{sec:relationships}.  
In \Cref{sec:tradeoff_RD}, we derive the rate-distortion functions for different permutation spaces. In
\Cref{sec:codes}, we provide achievability schemes for different permutation spaces under different regimes.
After that, we turn our attention to non-uniform distributional model over the permutation space and analyze the
lossless and lossy compression for Mallows model in \Cref{sec:mallows}. 
We conclude with a few remarks in \Cref{sec:conclu}.

\section{Problem formulation}
\label{sec:rd_formulation}
In this section we discuss aspects of the formulation of the rate-distortion problem for permutation spaces.
We first introduce the distortion measures of interest in \Cref{sec:distortion_defs}, 
and then provide a mathematical formulation of the rate-distortion problem in~\Cref{sec:rd_problem_def}.

\subsection{Notation and facts}
\label{sec:rd_notations}
Let $\PermutationSet$ denote the symmetric group of $n$ elements. We write an
element of $\PermutationSet$ as an array of natural numbers with values
ranging from $1, \dots, n$ and every value occurring only once in the array. For example,
$\sigma = [3,4,1,2,5]  \in \mathcal{S}_5$. 
This is also known as the {\em vector notation} for permutations. The identity of the symmetric group $\PermutationSet$ (identity permutation) is
denoted by $\idperm = [1, 2,\dots,n]$.
For a permutation $\sigma$, we denote its permutation inverse by $\sigma^{-1}$, where 
\begin{olequation*}
    \sigma^{-1}(x) = i \text{ when } \sigma(i) = x,
\end{olequation*}
and $\sigma(i)$ is the $i$-th element in array $\sigma$.
For example,
the permutation inverse of $\sigma = [2,5,4,3,1]$ is 
$\sigma^{-1} = [5, 1, 4, 3, 2]$.
Given a metric $d: \PermutationSet \times \PermutationSet \rightarrow \preals\cup \{0\}$, we
define a \emph{permutation space} $\PermutationSpace{d}$.

Throughout the paper, we let $[a:b] \defeq \{a,a+1,\dots, b-1, b\}$ for any two integers $a$ and $b$, and
use $\sigma[a:b]$ as a shorthand for the vector $[\sigma(a), \sigma(a+1), \ldots, \sigma(b)]$.

We make use of the following version of \emph{Stirling's approximation}:
\begin{equation}
	\label{eq:stirling_approx}
	\left(\frac{m}{e}\right)^m e^{\frac{1}{12m + 1}}
	<
	\frac{m!}{\sqrt{2\pi m}}
	<
	\left(\frac{m}{e}\right)^m e^{\frac{1}{12m}}
    , 
    m \geq 1
	.
\end{equation}

\subsection{Distortion measures}
\label{sec:distortion_defs}
There exists many natural distortion measures on the
permutation group $\PermutationSet$~\cite{deza_metrics_1998}. 
In this paper we choose a few distortion measures of interest in a variety
of application settings, including 
Spearman's footrule ($\ell_1$ distance between two permutation vectors), 
Chebyshev distance ($\ell_\infty$ distance between two permutation vectors), 
\ktdis\ and the \invLoneName~(see \Cref{def:invLone}).

Before introducing definitions for these distortion measures, 
we define the concept of \emph{ranking}.
Given a list of items with values $v_1, v_2, \ldots, v_n$ such that 
$v_{\invperm(1)} \succ v_{\invperm(2)} \succ \ldots \succ v_{\invperm(n)}$, 
where $a \succ b$ indicates
$a$ is preferred to $b$,  we say the permutation $\sigma$ is the \emph{ranking} of this
list of items, where $\sigma(i)$ provides the rank of item $i$, and $\invperm(r)$
provides the index of the item with rank $r$.
Note that sorting via pairwise comparisons is simply the procedure of rearranging 
$v_1, v_2, \ldots, v_n$ to 
$v_{\invperm(1)}, v_{\invperm(2)}, \ldots, v_{\invperm(n)}$ 
based on preferences obtained from pairwise comparisons.

Given two rankings $\sigma_1$ and $\sigma_2$, we measure the total deviation of ranking and maximum deviation of
ranking by the {Spearman's footrule} and the {Chebyshev distance} respectively.
\begin{definition}[Spearman's footrule~\cite{diaconis_spearmans_1977}]
\label{def:spearman_footrule}
Given two permutations $\sigma_1, \sigma_2 \in \PermutationSet$, 
the \emph{Spearman's footrule} between $\sigma_1$ and $\sigma_2$ is
\begin{equation*}
    \dLone{\sigma_1}{\sigma_2} \defeq \onenorm{\sigma_1 - \sigma_2} 
    = \sum_{i=1}^n \abs{\sigma_1(i) - \sigma_2(i)}
    .
\end{equation*}
\end{definition}
\begin{definition}[Chebyshev distance]
Given two permutations $\sigma_1, \sigma_2 \in \PermutationSet$, 
the \emph{Chebyshev distance} between $\sigma_1$ and $\sigma_2$ is
\begin{equation*}
    \dLinf{\sigma_1}{\sigma_2} \defeq \infnorm{\sigma_1 - \sigma_2} 
    = \max_{1 \leq i \leq n} \abs{\sigma_1(i) - \sigma_2(i)}
    .
\end{equation*}
\end{definition}
The Spearman's footrule in $\PermutationSet$ is upper bounded by $\floor{n^2/2}$ (\cf\
\Cref{tab:asy_char}) and the Chebyshev distance in $\PermutationSet$ is upper bounded by $n-1$.

Given two lists of items with ranking $\sigma_1$ and $\sigma_2$, 
let $\pi_1 \defeq \invperm_1$ and $\pi_2 \defeq \invperm_2$, 
then we define the number of pairwise adjacent swaps on $\pi_1$ that changes the ranking of
$\pi_1$ to the ranking of $\pi_2$ as the \ktdis.
\begin{definition}[\ktdis~\cite{kendall_new_1938}]
    The \emph{\ktdis} $d_{\tau}(\sigma_1, \sigma_2)$ from one
    permutation $\sigma_1$ to another permutation $\sigma_2$ is defined as the
    minimum number of transpositions of pairwise adjacent elements required to
    change $\sigma_1$ into $\sigma_2$.
\end{definition}
The \ktdis{} is upper bounded by $\nchoosek{n}{2}$.

\begin{example}[\ktdis]
    The \ktdis{} for $\sigma_1 = [1,5,4,2,3]$ and $\sigma_2 = [3,4,5,1,2]$ is 
    $\dtau{\sigma_1}{\sigma_2} = 7$, as one needs at least 7 transpositions of pairwise
    adjacent elements to change $\sigma_1$ to $\sigma_2$. For example, 
    \begin{align*}
        \sigma_1 &= [1,5,4,\mathbf{2,3}] 
        \\
        &\rightarrow [1,5,\mathbf{4,3},2] \rightarrow [1,\mathbf{5,3},4,2] \rightarrow [\mathbf{1,3},5,4,2] 
        \\
        &\rightarrow [3,\mathbf{1,5},4,2] \rightarrow [3,5,\mathbf{1,4},2] \rightarrow [3,\mathbf{5,4},1,2] 
        \\
        &\rightarrow [3,4,5,1,2] = \sigma_2
        .
    \end{align*}
\end{example}
Being a popular global measure of disarray in statistics, 
\ktdis{} also has a natural connection to sorting algorithms. In particular, 
given a list of items with values $v_1, v_2, \ldots, v_n$ such that
$v_{\invperm(1)} \succ v_{\invperm(2)} \succ \ldots \succ v_{\invperm(n)}$, 
$\dtau{\invperm[\sigma]}{\idperm}$ is the number of swaps needed to
sort this list of items in a bubble-sort algorithm~\cite{knuth_art_1998}.

Finally, we introduce a distortion measure based on the concept of inversion vector, another measure of the
order-ness of a permutation. 
\begin{definition}[inversion, inversion vector~\cite{knuth_art_1973}]
    \label{def:inversion}
    An \emph{{inversion}} in a permutation $\sigma \in \PermutationSet$ is a pair
    $(\sigma(i), \sigma(j))$ 
    such that $i < j$ and $\sigma(i) > \sigma(j)$.

    We
    use $I_n(\sigma)$ to denote the total number of inversions in $\sigma \in \PermutationSet$, 
    and 
    \begin{equation}
    \label{eq:K_nk}
    K_n(k) \defeq \cardinality{ \SetDef{\sigma \in \PermutationSet}{I_n(\sigma) = k} }
    \end{equation}
    to denote the number of permutations with $k$ inversions.

    Denote $i' = \sigma(i)$ and $j' = \sigma(j)$, 
    then 
    $i = \invperm(i')$ and $j = \invperm(j')$,  
    and thus 
    $i < j$ and $\sigma(i) > \sigma(j)$ is equivalent to
    $\invperm(i') < \invperm(j')$ and $i' > j'$. 

    A permutation $\sigma \in \PermutationSet$ is associated with an \emph{{inversion vector}}
    $\inv{\sigma} \in \InvVectorSet \defeq [0:1] \times [0:2] \times \cdots \times [0:n-1]$,
    where $\inv{\sigma}(i'), 1 \leq i' \leq n-1$ is the number of inversions in $\sigma$ in
    which $i'+1$ is the first element.
    Formally, for $i' = 2, \ldots, n$, 
    \begin{equation*}
        \label{eq:inv_vector_def}
        \inv{\sigma}(i'-1) = \cardinality{
        \SetDef{j' \in [1:n]}{j' < i', \invperm(j') > \invperm(i') }
        }
        .
    \end{equation*}
\end{definition}
Let $\pi \defeq \invperm$, then the \emph{inversion vector} of $\pi$, $\inv{\pi}$,
measures the deviation of ranking $\sigma$ from $\idperm$. In particular, 
note that 
\begin{align*}
\invx{\pi}{k} 
&= 
\cardinality{\SetDef{j' \in [1:n]}{j' < k, \invperm[\pi](j') > \invperm[\pi](k) }}
\\
&= \cardinality{\SetDef{j' \in [1:n]}{j' < k, \sigma(j') > \sigma(k) }}
\end{align*}
indicates the number of elements that have \emph{larger ranks}
and \emph{smaller item indices} than that of the element with index $k$.  
In particular, the
rank of the element with index $n$ is $n - \invx{\pi}{n-1}$. 
\begin{example}
    Given 5 items such that
    \begin{olequation*}
        v_4 \succ v_1 \succ v_2 \succ v_5 \succ v_3
        ,
    \end{olequation*}
    then the inverse of the ranking permutation is $\pi = [4,1,2,5,3]$, 
    with inversion vector $\inv{\pi} = [0, 0, 3, 1]$. Therefore, the rank of the
    $v_5$ is $n-\invx{\pi}{n-1} = 5 - 1 = 4$.
\end{example}
The mapping from $\PermutationSet$ to $\InvVectorSet$ is one-to-one as the inversion vectors exactly describes
the execution of the algorithm insertion sort~\cite{knuth_art_1998}.

With these, we define the \invLoneName.
\begin{definition}[\invLoneName]
\label{def:invLone}
Given two permutations $\sigma_1, \sigma_2 \in \PermutationSet$, 
we define the {\invLoneName}, $\ell_1$ distance of two inversion vectors, as 
\begin{equation}
    \dinvLone{\sigma_1}{\sigma_2} 
    \defeq
    \sum_{i=1}^{n-1}|\inv{\sigma_1}(i)- \inv{\sigma_2}(i)|
    \label{eq:Loneinv_def}
    .
\end{equation}
\end{definition}

\begin{example}[\invLoneName]
    The inversion vector for permutation $\sigma_1 = [1,5,4,2,3]$ is 
    $\inv{\sigma_1} = [0, 0, 2, 3]$, as the inversions are 
    $(4, 2), (4, 3), (5, 4), (5, 2), (5, 3)$.
    The inversion vector for permutation $\sigma_2 = [3,4,5,1,2]$ is 
    $\inv{\sigma_2} = [0, 2, 2, 2]$, as the inversions are 
    $(3, 1), (3, 2), (4, 1), (4, 2), (5, 1), (5, 2)$.
    Therefore, 
    \begin{equation*}
        \dinvLone{\sigma_1}{\sigma_2} = \dLone{[0,0,2,3]}{[0,2,2,2]} = 3 
    .
    \end{equation*}
\end{example}
As we shall see in \Cref{sec:relationships}, all these distortion measures are related.  While the
operational significance of the \invLoneName\ may not be as clear as other distortion measures, some of its
properties provide useful insights in the analysis of other distortion measures.

\begin{remark}
While Spearman's footrule and Chebyshev distance operate on the \emph{ranking domain}, inversion
vector and Kendall tau distance can be viewed as operating on \emph{the inverse of the ranking domain}. 
\end{remark}

%

\subsection{Rate-distortion problems}
\label{sec:rd_problem_def}
With the distortions defined in \Cref{sec:distortion_defs}, in this section we define rate-distortion problems
under both average-case and worst-case distortions. 

\begin{definition}[Codebook for average-case distortion]
    \label{def:codebook_ave_case}
    An $(n, D_n)$ source code $\CodeA \allowbreak \subseteq \PermutationSet$ for $\PermutationSpace{d}$ under the average-case distortion is 
    a set of permutations such that
    for a $\sigma$ that is drawn from $\PermutationSet$ according to a distribution
    $P$ on $\PermutationSet$, 
    there exists an encoding mapping $f_n: \PermutationSet \rightarrow \CodeA$ that
    \begin{align}
        \label{eq:ave_d}
        \Ep{P}{d(f_n(\sigma), \sigma)} \leq D_n
        .
    \end{align}
    The mapping $f_n: \PermutationSet \rightarrow \CodeA$ can be assumed to satisfy
    \begin{equation*}
        f_n(\sigma) = \argmin_{\sigma' \in \CodeA} d(\sigma', \sigma)
    \end{equation*}
    for any $\sigma \in \PermutationSet$.
\end{definition}
In most parts of this paper we focus on the case $P$ is uniformly distributed over the symmetric group
$\PermutationSet$, except in \Cref{sec:mallows}, where a distribution arising from the Mallows model is used.  In
both cases the source distribution has support $\PermutationSet$, and we define the worst-case distortion as
follows.
\begin{definition}[Codebook for worst-case distortion]
    An $(n;D_n)$ source code $\CodeW \subseteq S_n$ for $\PermutationSpace{d}$ under the worst-case distortion is a set of
    permutations such that for any $\sigma \in \PermutationSet$, 
    there exists an encoding mapping $f_n: \PermutationSet \rightarrow \CodeA$ that
    \begin{align}
        \label{eq:worst_d}
        \max_{\sigma \in \PermutationSet}{d(f_n(\sigma), \sigma)} \leq D_n
        .
    \end{align}
    The mapping $f_n: \PermutationSet \rightarrow \CodeW$ can be assumed to satisfy
    \begin{equation*}
        f_n(\sigma) = \argmin_{\sigma' \in \CodeW} d(\sigma', \sigma)
    \end{equation*}
    for any $\sigma \in \PermutationSet$.
\end{definition}

\begin{definition}[Rate function]
For a class of source codes $\Set{\CodeG}$ that achieve a distortion $D_n$, 
let $\Ag(n, D_n)$ be the minimum size of such codes, and we define the minimal rate for distortions $D_n$ as 
\begin{equation*}
    \RateG(D_n) \defeq  \frac{\log A(n,D_n)}{\log n!}
    .
\end{equation*}
In particular, we denote the minimum rate of the codebook under average-case distortion with uniform source
distribution and worst-case distortions by $\RateA\left( D_n \right)$ and $\RateW\left( D_n \right)$
respectively.
\end{definition}

Similar to the classical rate-distortion setup, we are interested in deriving the
trade-off between distortion level $D_n$ and the rate $R(D_n)$ as $\ntoinf$. 
In this work we show that for the distortions $d(\cdot,\cdot)$ 
and the sequences of distortions $\Set{D_n, n \in \pints}$ of interest, $\lim_{\ntoinf} R(D_n)$ exists. 

For \ktdis\ and \invLoneName, a close observation shows that in regimes such as
$D_n = O(n)$ and $D_n = \BigTheta{n^2}$, $\lim_{\ntoinf} R(D_n)=1$ and
$\lim_{\ntoinf} R(D_n)=0$ respectively. In these two regimes, 
the trade-off between rate and distortion is really shown in the
higher order terms in $\log A(n,D_n)$, \ie, 
\begin{align}
    \label{eq:higher_order_term}
    r(D_n) \defeq \log A(n,D_n) - \log n! \lim_{\ntoinf} R(D_n) 
    .
\end{align}
For convenience, we categorize the distortion $D_n$ under \ktdis\ or \invLoneName\ into three regimes.  
We say $D$ is small when $D=\BigO{n}$, moderate when $D=\BigTheta{n^{1+\delta}}, 0 < \delta < 1$, and large
when $D=\BigTheta{n^2}$\footnote{%
In the small distortion region with $R(D_n) = 1$, $r(D_n)$ is negative while in the large
distortion region where $R(D_n) = 0$, $r(D_n)$ is positive.
}.

We choose to omit the higher order term analysis for $\PermutationSpace{\dLoneText}$ because its analysis is
essentially the same as $\PermutationSpace{\dtauText}$, and the analysis for $\PermutationSpace{\dLinfText}$ is
still open.

Note that the higher order terms $r(D_n)$ may behave differently under average and worst-case distortions, and
in this paper we restrict our attention to the worst-case distortion.

\section{Relationships between distortion measures}
\label{sec:relationships}
In this section we show how the four distortion measures defined in \Cref{sec:distortion_defs} are
related to each other, which is summarized in \eqref{eq:geq_rlships} and \eqref{eq:leq_rlships}.
These relationships imply equivalence in some lossy compression schemes, which we exploit 
to derive the rate-distortion functions in \Cref{sec:tradeoff_RD}.

For any $\sigma_1 \in \PermutationSet$ and $\sigma_2$ randomly uniformly chosen from $\PermutationSet$, the following relations hold:
\begin{alignat}{2}
    n \dLinf{\sigma_1}{\sigma_2}
    &\geq&&
    \dLone{\sigma_1}{\sigma_2}
    \nn
    \\
    &\geq&&
    \dtau{\invperm_1}{\invperm_2}
    \nn
    \\
    &\geq&&
    \dinvLone{\invperm_1}{\invperm_2}
    ,
    \label{eq:geq_rlships}
    \\
    n \dLinf{\sigma_1}{\sigma_2}
    &\stackrel{\whp}{\lesim}&&
    \dLone{\sigma_1}{\sigma_2}
    \nn
    \\
    &\,\;\;{\lesim}\;&&
    \dtau{\invperm_1}{\invperm_2}
    \nn
    \\
    &\stackrel{\whp}{\lesim}&&
    \dinvLone{\invperm_1}{\invperm_2}
    ,
    \label{eq:leq_rlships}
\end{alignat}
where $x \lesim y$ indicates $x < c \cdot y$ for some constant $c > 0$, 
and $\stackrel{\whp}{\lesim}$ indicates $\lesim$ with high probability.

Next, we provide detailed arguments for \eqref{eq:geq_rlships} and \eqref{eq:leq_rlships} by analyzing the relationship
between different pairs of distortion measures.

\subsubsection{Spearman's footrule and Chebyshev distance}
Let $\sigma_1$ and $\sigma_2$ be any permutations in $\PermutationSet$, then by definition,
\begin{equation}
    \dLone{\sigma_1}{\sigma_2} \leq n \cdot \dLinf{\sigma_1}{\sigma_2}, 
    \label{eq:inf_ub_l1}
\end{equation}
and additionally, a scaled Chebyshev distance lower bounds the Spearman's footrule with high
probability. More specifically, 
for any $\pi \in \PermutationSet$, let $\sigma$ be a permutation chosen uniformly from
$\PermutationSet$, then
\begin{equation}
    \Prob{ c_1 \cdot n \cdot \dLinf{\pi}{\sigma} \leq \dLone{\pi}{\sigma} }
    \geq 1 - \BigO{1/n}
    \label{eq:linf_lb_lone_prob}
\end{equation}
for any positive constant $c_1 < 1/3$ (See \Cref{sec:proof_linf_lb_lone_prob} for proof).

\subsubsection{Spearman's footrule and \ktdis}
The following theorem 
is a well-known result on 
the relationship between the \ktdis{} and the $\ell_1$ distance of permutation vectors.
\begin{theorem}[\cite{diaconis_spearmans_1977}]
    \label{thm:l1_and_kt}
Let $\sigma_1$ and $\sigma_2$ be any permutations in $\PermutationSet$, then
\begin{equation}
    d_{\ell_1}(\sigma_1, \sigma_2)/2
    \leq 
    d_{\tau}(\sigma_1^{-1}, \sigma_2^{-1}) 
    \leq
    d_{\ell_1}(\sigma_1, \sigma_2) 
    \label{eq:kt_l1_dis}
    .
\end{equation}
\end{theorem}

\subsubsection{\invLoneName\ and \ktdis}
We show that the \invLoneName\ and the \ktdis{} are
related via \Cref{thm:kt_lb_l1}.
\begin{theorem} 
    \label{thm:kt_lb_l1}
    Let $\sigma_1$ and $\sigma_2$ be any permutations in $\PermutationSet$, then for $n \geq 2$, 
    \begin{equation}
        \frac{1}{n-1} \dtau{\sigma_1}{\sigma_2} 
        \leq \dinvLone{\inv{\sigma_1}}{\inv{\sigma_2}}
        \leq d_{\tau}(\sigma_1, \sigma_2)
        .
        \label{eq:kt_bounds_l1inv}
    \end{equation}
\end{theorem} 
\begin{IEEEproof}
    See \Cref{sec:proof_kt_bounds_l1}.
\end{IEEEproof}

\begin{remark}
    The lower and upper bounds in \Cref{thm:kt_lb_l1} are tight in the sense that there exist permutations $\sigma_1$ and $\sigma_2$ that
    satisfy the equality in either lower or upper bound.
    For equality in lower bound, when $n = 2m$, let 
    $\sigma_1 = [1, 3, 5, \ldots, 2m-3, 2m-1, 2m, 2m-2, \ldots, 6, 4, 2]$,
    $\sigma_2 = [2, 4, 6, \ldots, 2m-2, 2m, 2m-1, 2m-3, \ldots, 5, 3, 1]$,
    then $\dtau{\sigma_1}{\sigma_2} = n(n-1)/2$ and $\dinvLone{\sigma_1}{\sigma_2} = n/2$, as
    $\inv{\sigma_1} = [0, 0, 1, 1, 2, 2, \ldots, m-2, m-2, m-1, m-1]$, 
    $\inv{\sigma_2} = [0, 1, 1, 2, 2, 3, \ldots, m-2, m-1, m-1, m]$.
    For equality in upper bound, note that $\dtau{\idperm}{\sigma} = \dinvLone{\idperm}{\sigma}$.
\end{remark}

\Cref{thm:kt_lb_l1} shows that  in general  $\dtau{\sigma_1}{\sigma_2}$ is not a good
approximation to $\dinvLone{\sigma_1}{\sigma_2}$ due to the $1/(n-1)$ factor.  However,
\eqref{eq:kt_lb_l1inv_prob} shows that \ktdis{} scaled by a constant actually provides a lower
bound to the \invLoneName\ with high probability.
In particular, for any $\pi \in \PermutationSet$, 
let $\sigma$ be a permutation chosen uniformly from $\PermutationSet$, then
\begin{equation}
    \Prob{ c_2 \cdot \dtau{\pi}{\sigma} \leq \dinvLone{\pi}{\sigma} }
    \geq 1 - \BigO{1/n}
    \label{eq:kt_lb_l1inv_prob}
\end{equation}
for any positive constant $c_2 < 1/2$ (See \Cref{sec:proof_kt_bounds_l1_prob} for proof).


Results in both \eqref{eq:linf_lb_lone_prob} and \eqref{eq:kt_lb_l1inv_prob} are concentration results in the
sense that the mean for distances are $\BigTheta{n^2}$ and the standard deviation for the distances are
$\BigTheta{n^{3/2}}$. Related quantities are summarized in \Cref{tab:asy_char}, where 
results on $\LoneText$ distance and \ktdis\ are from \cite[Table 1]{diaconis_spearmans_1977}, 
and results on $\LinfText$ distance and
\invLoneName\ are derived in \Cref{sec:proof_linf_lb_lone_prob} and \Cref{sec:proof_kt_bounds_l1_prob}.
Therefore, these distance are concentrated around mean and separated probabilistically.
\begin{table}
\centering
\caption{Characterization of maximum, mean and variance of various distances.}
\label{tab:asy_char}
\begin{tabular}[bt]{c|ccc}
& Max & Mean & Variance
\vspace{1pt}
\\
\hline
\vspace{1pt}
$n \cdot \LinfText$ & $n(n-1)$ & $< n^2$ & $\BigTheta{n^3}$
\\ 
$\LoneText$ & $\floor{n^2/2}$ & $n^2/3 + o({n^2})$ & $2n^3/45 + o(n^3)$
\\
Kendall-tau & $n(n-1)/2$ & $n^2/4 + o(n^2)$ & $n^3/36 + o(n^3)$
\\
inversion-$\ell_1$ & $n(n-1)/2$ & $> n^2/8$ & $< n^3/6$
\\
\hline
\end{tabular}
\vspace{1ex}
\end{table}

\begin{remark}
The constants in \eqref{eq:linf_lb_lone_prob} and \eqref{eq:kt_lb_l1inv_prob} may be improved if both of the
permutations in question are chosen randomly, instead of one being fixed. However as the techniques are exactly
same, we refrain from providing those expressions.
\end{remark}

\section{Trade-offs between rate and distortion}
\label{sec:tradeoff_RD}
In this section we present some of the  main results of this paper---the trade-offs between rate and distortion in
permutation spaces. Throughout this section we assume the permutations are uniformly distributed over
$\PermutationSet$.

We first present \Cref{thm:RD_equivalence}, which shows how a lossy source code under one
distortion measure implies a lossy source code under another distortion measure. 
Building on these relationships, \Cref{thm:RD_funcs} shows that all distortion measures in this
paper essentially share the same rate-distortion function.
Last, in \Cref{sec:higher_order_term}, we present results on the trade-off between rate and
distortion for $\PermutationSpace{\dtauText}$ and $\PermutationSpace{\dinvLoneText}$
when the distortion leads to degenerate rates $R(D_n) = 0$ and $R(D_n) = 1$.

\subsection{Rate-distortion functions}
\label{sec:rdfs}
\begin{figure}[!t]
    \centering
    \addplot{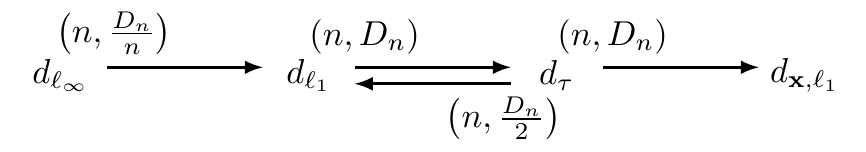}
    \caption{Relationship between source codes. An arrow indicates
        a source code in one space implies a source in another space.
    }
    \label{fig:equivalence_diagram}
\end{figure}

\begin{theorem}[Relationships of lossy source codes]
    \label{thm:RD_equivalence}
    For both \emph{worst-case distortion} and \emph{average-case distortion with uniform distribution}, a
    following source code on the left hand side implies a source code on the right hand side:
    \begin{enumerate}
        \item $(n, D_n/n)$ source code for $\PermutationSpace{\dLinfText}$ 
            $\Rightarrow$ $(n, D_n)$ source code for $\PermutationSpace{\dLoneText}$, 
        \item $(n, D_n)$ source code for $\PermutationSpace{\dLoneText}$
            $\Rightarrow$ $(n, D_n)$ source code for $\PermutationSpace{\dtauText}$,
        \item $(n, D_n)$ source code for $\PermutationSpace{\dtauText}$ 
            $\Rightarrow$ $(n, 2 D_n)$ source code for $\PermutationSpace{\dLoneText}$, 
        \item $(n, D_n)$ source code for $\PermutationSpace{\dtauText}$ 
            $\Rightarrow$ $(n, D_n)$ source code for $\PermutationSpace{\dinvLoneText}$.
    \end{enumerate}
\end{theorem}
The relationship between source codes is summarized in \Cref{fig:equivalence_diagram}. 

\begin{remark}[Non-equivalence of lossy source codes for $\PermutationSpace{\dLoneText}$ and $\PermutationSpace{\dLinfText}$]
It is worth noting that in general, an $(n, D_n)$ source code for $\PermutationSpace{\dLoneText}$  does not imply
an $(n, D_n/(nc_1) + \BigO{1})$ source code for $\PermutationSpace{\dLinfText}$ in spite of the relationship
shown in \eqref{eq:leq_rlships}, even under the \emph{average-case distortion}. 
This is exemplified in \Cref{eg:counter} below.

In~\cite{wang_lossy_2014}, it was shown incorrectly that lossy source codes for $\PermutationSpace{\dLoneText}$ and $\PermutationSpace{\dLinfText}$
are equivalent, leading to an over-generalized version of \Cref{thm:RD_equivalence}.
\end{remark}
\begin{example}
\label{eg:counter}
When $n = k m$ and $m = n^{\delta}$, we define the following $k$ sets with size $m$
\begin{align*}
\cI_1 &= \Set{2, 3, \ldots, m, m+1}, \\
\cI_2 &= \Set{m+2, m+3, \ldots, 2m, 2m+1}, \\
&\ldots\\
\cI_j &= \Set{(j-1)m+2, (j-1)m+3, \ldots, jm, jm+1}, \\
&\ldots\\
\cI_k &= \Set{(k-1)m+2, (k-1)m+3, \ldots, n, 1}
\end{align*}
and construct the following $k$ subsequences for any permutation $\sigma \in \PermutationSet$:
\begin{align*}
\mathbf{s}_j(\sigma) &= [\sigma(j_1), \sigma({j_2}), \ldots, \sigma({j_m})], 1 \leq j \leq k
\end{align*}
where for each $j$, $j_p \in \cI_j$ for any $1 \leq p \leq m$ and $j_1 < j_2 \cdots < j_m$.

Given any permutation $\sigma$, we can encode it as as $\hat{\sigma}$ by sorting each of its subsequences $\mathbf{s}_j(\sigma), 1 \leq j \leq k$. Then the 
overall distortion $\PermutationSpace{\dLoneText}$ satisfies
\begin{align*}
    D_{\LoneText} \leq (k-1) m^2/2 + \left[(m-1)^2 + 2n\right] &= \BigO{k m^2/2} 
    \\
    &= \BigO{n^{1+\delta}}.
\end{align*}
Therefore, this source code is an $(n, D_n)$ source code for $\PermutationSpace{\dLoneText}$. However, 
for any $\sigma$ in $\PermutationSet$, if $\sigma(1) \neq 1$, 
\begin{align*}
\dLinf{\sigma}{\hsigma} \geq (k-1)m+2-1 \geq (k-1)m = \BigTheta{n}.
\end{align*}
Hence this encoding achieves average distortion $\BigTheta{n}$ in $\PermutationSpace{\dLinfText}$. 
Therefore, while this code is $D_n$ for $\PermutationSpace{\dLoneText}$, it is not $D_n/n$ for $\PermutationSpace{\dLinfText}$. 

Similarly, one can find a code that achieves distortion $\BigO{n^{1+\delta}}$ for $\PermutationSpace{\dinvLoneText}$ but not $\PermutationSpace{\dtauText}$. 
\end{example}

The proof of \Cref{thm:RD_equivalence} is based on the relationships between various distortion measures investigated in
\Cref{sec:relationships} and we defer the proof details in \Cref{sec:proof_RD_equivalence}.

Below shows that, for the uniform distribution on $\PermutationSet$, the rate-distortion function is the same
for both average- and worst-case, apart from the terms that are asymptotically negligible.  

\begin{figure}[!t]
    \centering
        \addplot{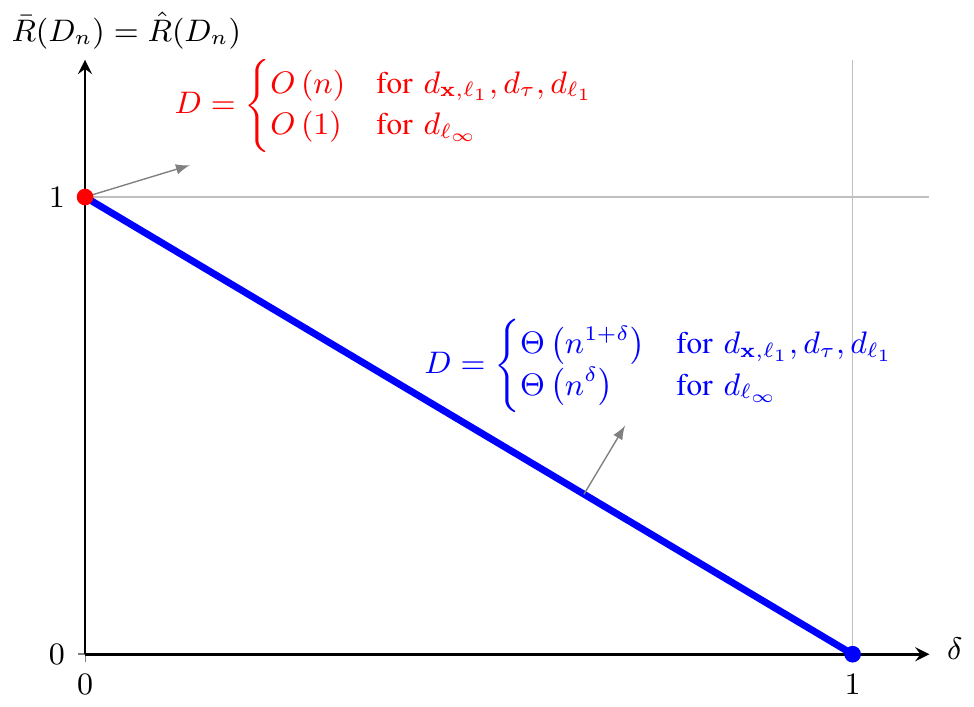}
    \caption{Rate-distortion function for permutation spaces 
    $\PermutationSpace{\dinvLoneText}$, $\PermutationSpace{\dtauText}$, $\PermutationSpace{\dLoneText}$, 
    and $\PermutationSpace{\dLinfText}$.}
    \label{fig:rdf_all}
\end{figure}

\begin{theorem}[Rate-distortion functions]
    \label{thm:RD_funcs}
    For permutation spaces $\PermutationSpace{\dinvLoneText}$, $\PermutationSpace{\dtauText}$,
    and $\PermutationSpace{\dLoneText}$, 
    \begin{align*}
        \label{eq:RD_funcs}
        \RateA(D_n) &= \RateW(D_n) 
        \\
        &= \begin{cases}
            1 & \text{if } D_n = \BigO{n},
            \\
            1 - \delta & \text{if } D_n = \BigTheta{n^{1+\delta}}, \quad 0 < \delta \leq 1.
        \end{cases}
        \nn
    \end{align*}
    For the permutation space $\PermutationSpace{\dLinfText}$, 
    \begin{align}
        \RateA(D_n) &= \RateW(D_n) 
        \\
        &= \begin{cases}
            1 & \text{if } D_n = \BigO{1},
            \\
            1 - \delta & \text{if } D_n = \BigTheta{n^{\delta}}, \quad 0 < \delta \leq 1 .
        \end{cases}
        \nn
        \label{eq:RD_func_linf}
    \end{align}
\end{theorem}
The rate-distortion functions for all these spaces are summarized in \Cref{fig:rdf_all}.
\begin{proof}[Proof sketch]
The achievability comes from the compression schemes\footnote{Achievability results can also follow from simple random choice construction of
covering codes,which are quite standard \cite{cohen_covering_1997}. Instead we provide explicit constructions.} proposed in \Cref{sec:codes}. 
The average-case converse for $\PermutationSpace{\dinvLoneText}$ can be shown via the geometry of permutation spaces in
\Cref{sec:geometry}. Then because a $D$-ball in $\PermutationSpace{\dinvLoneText}$ has the largest volume (\cf\
\eqref{eq:geq_rlships}), a converse for other permutation spaces can be inferred.

The rest of the proof follows from the simple fact that an achievability scheme for
the worst-case distortion is also an achievability scheme for the average-case
distortion, and a converse for the average-case distortion is also a converse for the
worst-case distortion.

We present the detailed proof in \Cref{sec:proof_RD_funcs}.
\end{proof}

Because the rate-distortion functions under average-case and worst-case distortion coincide,
if we require
\begin{equation}
    \lim_{\ntoinf} \Prob{d(f_n(\sigma), \sigma) > D_n} = 0
\end{equation}
instead of $\E{d(f_n(\sigma), \sigma)} \leq D_n$ in \Cref{def:codebook_ave_case}, 
then the asymptotic rate-distortion trade-off remains the same.

Given the number of elements $n$ and a distortion level $D$, we can compute the number of bits needed by first computing 
$\delta$ via the asymptotic relationship 
$\log D/\log n - 1$ 
(for permutation spaces $\PermutationSpace{\dinvLoneText}$, $\PermutationSpace{\dtauText}$, and $\PermutationSpace{\dLoneText}$)
or 
$\log D/\log n$ 
(for permutation space $\PermutationSpace{\dLinfText}$), then obtain the number of bits needed via $(1-\delta) n \log_2 n$.


%

%

\subsection{Higher order term analysis}
\label{sec:higher_order_term}
As mentioned in \Cref{sec:rd_formulation}, for small- and large-distortion regimes it is of interest to
understand the trade-off between rate and distortion via the higher order term defined in \eqref{eq:higher_order_term}. 
In this section  we present the analysis for both regimes in permutation spaces
$\PermutationSpace{\dtauText}$ and $\PermutationSpace{\dinvLoneText}$.
\begin{theorem}
    \label{thm:kt_RD_higher}
    In the permutation space $\PermutationSpace{\dtauText}$, 
    when $D_n = a n^\delta, 0 < \delta \leq 1$, for the worst-case distortion, 
    \begin{olequation*}
        \underline{r^{\sm}_{\ktText}(D_n)} \leq r(D_n) \leq \overline{r^{\sm}_{\ktText}(D_n)}
        ,
    \end{olequation*}
    where
    \begin{align}
        \underline{r^{\sm}_{\ktText}(D_n)} 
        &=
        \begin{cases}
            -a ( 1 - \delta) n^\delta \log n + \BigO{ n^\delta }
            ,\;
            0 < \delta < 1
        \\
            -n \left[ \log \frac{(1+a)^{1+a}}{a^a} \right] + \SmallO{n}
            ,\; 
            \delta = 1
        \end{cases}
        \hspace{-2ex}
        ,
        \label{eq:r_kt_lb}
        \\
        \overline{r^{\sm}_{\ktText}(D_n)}
        &=
        \begin{cases}
            - n^\delta \frac{a \log 2}{2} + \BigO{1} 
            ,\;
            0 < a < 1
            \\
            - n^\delta \frac{\log \floor{2a}!}{\floor{2a}} + \BigO{1}
            ,\;
            a \geq 1
        \end{cases}
        \hspace{-1ex}
        .
        \label{eq:r_kt_ub}
    \end{align}
    When $D_n = b n^2, 0 < b \leq 1/2$,
    \begin{olequation*}
        \underline{r^{\la}_{\ktText}(D_n)}
        \leq
        r(D_n) 
        \leq 
        \overline{r^{\la}_{\ktText}(D_n)}
        ,
    \end{olequation*}
    where
    \begin{align}
        \underline{r^{\la}_{\ktText}(D_n)}
        &=
        \max\Set{0, n \log \olbfrac{1}{2be^2}}, 
        \label{eq:r_kt_la_lb}
        \\
        \overline{r^{\la}_{\ktText}(D_n)}
        &=
        n \log \ceil{\olbfrac{1}{2b}} + \BigO{\log n}
        \label{eq:r_kt_la_ub}
        .
    \end{align}
\end{theorem}

\begin{remark}
Some of the results above for $\PermutationSpace{\dtauText}$, since their first appearances in the conference
version \cite{wang_rate-distortion_2013}, have been improved subsequently by
\cite{farnoud_rate-distortion_2014}. 
More specifically, for the small distortion regime, 
\cite[Lemma 7, Lemma 10]{farnoud_rate-distortion_2014} provides an improved upper bound and
show that $r^{\sm}_{\ktText}(D_n) = \underline{r^{\sm}_{\ktText}(D_n)}$ in \eqref{eq:r_kt_lb}.
For the large distortion regime, 
\cite[Lemma 11]{farnoud_rate-distortion_2014} shows a lower bound that is tighter than \eqref{eq:r_kt_la_lb}.
\end{remark}

\begin{theorem}
    In the permutation space $\PermutationSpace{\dinvLoneText}$, 
    when $D_n = a n^\delta, 0 < \delta \leq 1$, 
    \begin{equation*}
        \underline{r^{\sm}_{\invLoneText}(D_n)} 
        \leq r(D_n) \leq 
        \overline{r^{\sm}_{\invLoneText}(D_n)}
        ,
    \end{equation*}
    where $\underline{r^{\sm}_{\invLoneText}(D_n)} = \underline{r^{\sm}_{\ktText}(D_n)} - n^\delta
    \log 2$ (\cf\ \eqref{eq:r_kt_lb})
    and
    \begin{equation*}
        \overline{r_{\invLoneText}(D_n)}
        = 
        \begin{cases}
            - \floor{n^\delta} \log(2a - 1)
            & a > 1
            \\
            - \ceil{a n^\delta} \log 3
            & 0 < a \leq 1
        \end{cases}
        \label{eq:r_invdis_lb}
        .
    \end{equation*}
    When $D_n = b n^2, 0 < b \leq 1/2$,
    \begin{equation*}
        \underline{r^{\la}_{\invLoneText}(D_n)}
        \leq
        r(D_n) 
        \leq 
        \overline{r^{\la}_{\invLoneText}(D_n)}
        ,
    \end{equation*}
    where
    \begin{olequation*}
        \underline{r^{\la}_{\invLoneText}(D_n)} = 
        \underline{r^{\la}_{\ktText}(D_n)}
    \end{olequation*}
    (\cf\ \eqref{eq:r_kt_la_lb})
    and
    \begin{olequation*}
        \overline{r^{\la}_{\invLoneText}(D_n)}
        =
        n \log \ceil{\olbfrac{1}{4b}} + \BigO{1}
        \label{eq:r_inv_la_ub}
        .
    \end{olequation*}
    \label{thm:invdis_RD_higher}
\end{theorem}
\begin{IEEEproof}[Proof for \Cref{thm:kt_RD_higher} and \Cref{thm:invdis_RD_higher}]
    The achievability is presented in \Cref{sec:compress_small} and
    \Cref{sec:compress_large}. For converse, note that for a distortion measure $d$,
    \begin{equation*}
        \cardinality{\cC_n} N_d(D_n) \geq n!
        ,
    \end{equation*}
    where $N_d(D_n)$ is the maximum size of balls with radius $D_n$ in the
    corresponding permutation space $\PermutationSpace{d}$~(\cf\ \Cref{sec:geometry} for definitions), 
    then a lower bound on $\cardinality{\cC_n}$ follows from the upper bound on
    $N_d(D_n)$ in \Cref{lemma:logT_smallD} and \Cref{lemma:logT_largeD}. We omit the
    details as it is analogous to the proof of \Cref{thm:RD_funcs}.
\end{IEEEproof}

\begin{figure}[tb]
    \begin{center}
            \addplot{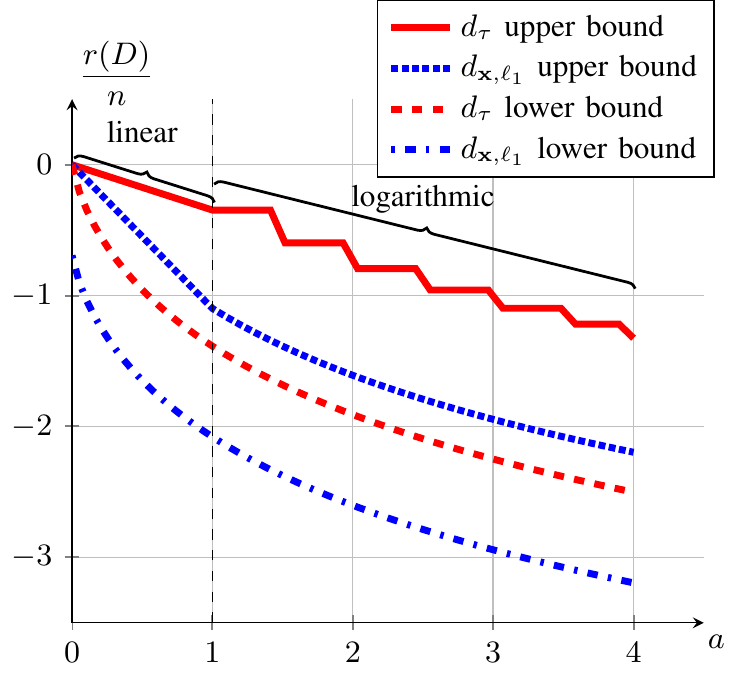}
    \end{center}
    \caption{
        Higher-order trade-off between rate and distortion in the small distortion
        regime with $D=an$. 
        The slope discontinuities of the $\dtauText$ upper bound in the range of $a \geq 1$ is due to the flooring in \eqref{eq:r_kt_ub}.
    }
    \label{fig:small_dis}
\end{figure}

\begin{figure}[tb]
    \begin{center}
        \addplot{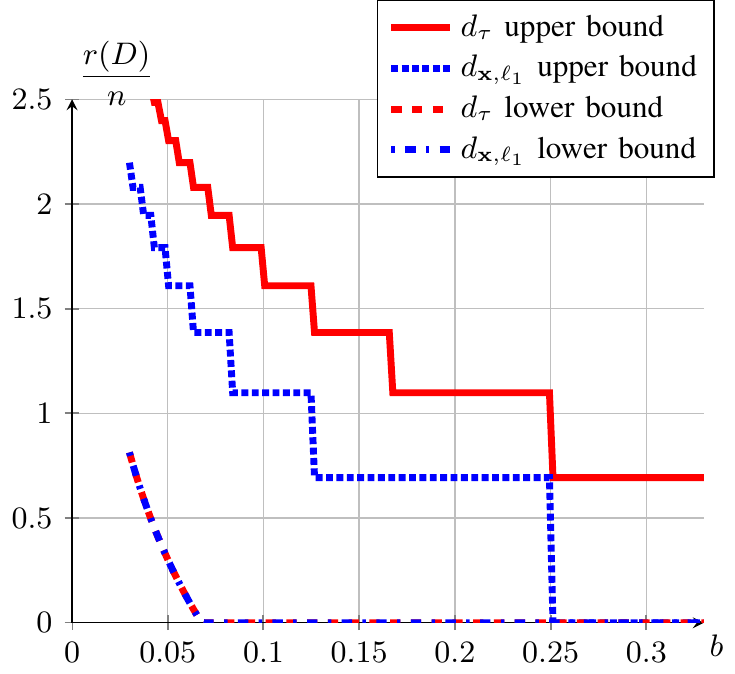}
    \end{center}
    \caption{
        Higher-order trade-off between rate and distortion in the large distortion
        regime with $D=bn^2$.
        The lower bounds for $\dtauText$ and $\dinvLoneText$ are identical.
    }
    \label{fig:large_dis}
\end{figure}

The bounds to $r(D_n)$ of both \ktdis\ and \invLoneName\ in both small
and large distortion regimes are shown in \Cref{fig:small_dis} and \Cref{fig:large_dis}.


\section{Compression schemes}
\label{sec:codes}
Though the permutation space has a complicated structure, in this section we show two rather straightforward
compression schemes, \emph{sorting subsequences} and \emph{component-wise scalar quantization}, which are optimal as they achieve the
rate-distortion functions in \Cref{thm:RD_funcs}.
We first describe these two key compression schemes in 
\Cref{sec:sort_to_quantize} and \Cref{sec:scalar_quantization} respectively.  
Then in \Cref{sec:compress_small,sec:compress_moderate,sec:compress_large}, we show that by simply applying these
schemes with proper parameters, we can achieve the corresponding trade-offs between rate
and distortion shown in \Cref{sec:tradeoff_RD}. 

The equivalence relationships in \Cref{thm:RD_equivalence} suggest these two compression schemes achieve the
same asymptotic performance. In addition, it is not hard to see that in general sorting subsequences has higher
time complexity (\eg, $\BigO{n \log n}$ for moderate distortion regime) than the time complexity of
component-wise scalar quantization (\eg, $\BigO{n}$ for moderate distortion regime).  However, these two
compression schemes operate on the permutation domain and the inversion vector of
permutation domain respectively, and the time complexity to convert a permutation from its vector representation
to its inversion vector representation is $\BigTheta{n \log n}$~\cite[Exercise 6 in Section
5.1.1]{knuth_art_1998}. Therefore, the cost of transforming a permutation between different representations should be taken
into account when selecting the compression scheme.

\subsection{Quantization by sorting subsequences}
\label{sec:sort_to_quantize}
In this section we describe the basic building block for lossy source coding in
permutation space $\PermutationSpace{\dLoneText}$, $\PermutationSpace{\dLinfText}$ and
$\PermutationSpace{\dtauText}$: sorting subsequences, either of the given permutation
$\sigma$ or of its inverse $\invperm$.
This operation reduces the number of possible permutations and thus the code rate,
but introduces distortion. 
By choosing the proper number of subsequences with proper
lengths, we can achieve the corresponding rate-distortion function.

\begin{figure}[b!]
    \centering
    \addplot{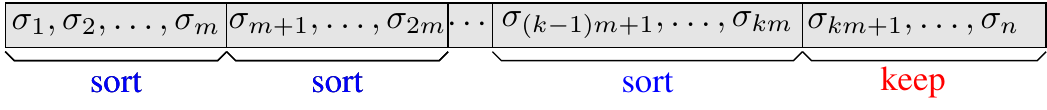}
    \caption{Quantization by sorting subsequences.}
    \label{fig:sorting_subsequences}
\end{figure}

More specifically, we consider 
a code obtained by the sorting the first $k$
subsequences with length $m$, $2 \leq m \leq n$, $k m \leq n$:
\begin{align*}
    \cC(k,m,n) 
    &\defeq \SetDef{f_{k,m}(\sigma)}{\sigma \in \PermutationSet}
\end{align*}
where $\sigma' = f_{k,m}(\sigma)$ satisfies
\begin{align*}
    &\sigma'[im+1:(i+1)m] 
    \\&\,\qquad = \fsort{\sigma[im+1:(i+1)m]}, & 0 \leq i \leq k ,
    \\
    &\sigma'(j) 
    ={} \sigma(j), & j > km 
    .
    \label{eq:quantize_by_sorting}
\end{align*}
This procedure is illustrated in \Cref{fig:sorting_subsequences}.

Then 
\begin{olequation*}
    \cardinality{\cC(k,m,n)}
    = 
    \olbfrac{n!}{m!^k}
    ,
\end{olequation*}
and we define the (log) size reduction as 
\begin{align*}
    \Delta(k,m) 
    &\defeq \log \frac{n!}{\cardinality{\cC(k,m,n)}}
    = k \log m!
    \\
    &\overset{(a)}{=} k \left[  m \log (m/e) + \frac{1}{2}{\log m} + \BigO{\frac{1}{m}}\right]
    ,
\end{align*}
where $(a)$ follows from Stirling's approximation in \eqref{eq:stirling_approx}.
Therefore, 
\begin{equation*}
    \Delta(k,m) = \begin{cases}
        k m \log m + \SmallO{km \log m} & m = \BigOmega{1}
        \\
        k \log m! & m = \BigTheta{1}
    \end{cases}
    .
\end{equation*}

We first calculate the worst-case and average-case distortions
for permutation space $\PermutationSpace{\dtauText}$:
\begin{align}
    \fDtauW{k,m} &= k \frac{m(m-1)}{2} \leq km^2/2
    \label{eq:DtauW}
    \\
    \fDtauA{k,m} &= k \frac{m(m-1)}{4} \leq km^2/4
    \label{eq:DtauA}
\end{align}
where \eqref{eq:DtauW} is from \eqref{eq:E_kt_dis}.
\begin{remark}
    \label{remark:kt_and_lone}
    Due to the close relationship between the \ktdis{} and the \spearman\
    shown in \eqref{eq:kt_l1_dis}, the following codebook 
    via the inverse permutations $\Set{\sigma^{-1}}$ 
    is an equivalent construction to the codebook for \ktdis\ above.
    \begin{enumerate}
        \item Construct a vector $a(\sigma)$ such that for $1 \leq i \leq k$,
            \begin{equation*}
                a(i) = j \text{ if } \sigma^{-1}(i) \in  [(j-1)m + 1, j m], 1 \leq j \leq k
                .
            \end{equation*}
            Then $a$ contains exactly $m$ values of integers $j$.
        \item Form a permutation $\pi'$ by replacing the length-$m$ subsequence
            of $a$ that corresponds to value $j$ by vector 
            $[(j-1)m + 1, (j-1)m + 2, \ldots, jm]$.
    \end{enumerate}
    It is not hard to see that the set of $\Set{\pi'^{-1}}$ forms a codebook
    with the same size with distortion in \ktdis\ upper bounded by $k m^2/2$.
\end{remark}

Similarly, for permutation space $\PermutationSpace{\dLoneText}$ and
$\PermutationSpace{\dLinfText}$, we consider sorting subsequences in the
\emph{inverse permutation} domain, where 
\begin{align*}
    \cC'(k,m,n) 
    &\defeq \SetDef{\invperm[\pi]}{\pi = f_{k,m}(\invperm), \sigma \in \PermutationSet}
    .
\end{align*}
It is straightforward that $\cC'(k,m,n)$ has the same cardinality as $\cC(k,m,n)$ and
hence code rate reduction $\Delta(k,m)$.
And the worst-case and average-case distortions satisfy
\begin{align}
    \fDLinfW{k,m} &= m-1 
    \\
    \fDLinfA{k,m} &\leq m -1 
    \\
    \fDLoneW{k,m} &= k\floor{m^2}/2 \leq k m^2/2 
    \label{eq:DLoneW}
    \\
    \fDLoneA{k,m} &= k (m^2-1)/3 
    \label{eq:DLoneA}
    ,
\end{align}
where \eqref{eq:DLoneW} comes from \Cref{tab:asy_char} and \eqref{eq:DLoneA}
comes from \eqref{eq:mean_l1}.

\subsection{Component-wise scalar quantization}
\label{sec:scalar_quantization}
To compress in the space of $\PermutationSpace{\dinvLoneText}$, component-wise scalar
quantization suffices, due to the product structure of the inversion vector space
$\InvVectorSet$.

More specifically, to quantize the $k$ points in $[0:k-1]$, where $k = 2, \cdots, n$, with $m$ uniformly spaced points, the maximal distortion is
\begin{equation}
    \fDinvLoneW{k,m} = \ceil{\left(\olfrac{k}{m} - 1\right)/2}
    ,
    \label{eq:D_from_m}
\end{equation}
Conversely, to achieve distortion $\DinvLoneW$ on $[0:k-1]$, we need 
\begin{equation}
    m = \ceil{\olbfrac{k}{2\DinvLoneW + 1}}
    \label{eq:m_from_D}
\end{equation}
points.

\subsection{Compression in the moderate distortion regime}
\label{sec:compress_moderate}
In this section we provide compression schemes in the moderate distortion regime, 
where for any $0 < \delta < 1$, $D_n = \BigTheta{n^{\delta}}$ for $\PermutationSpace{\dLinfText}$
and
$D_n = \BigTheta{n^{1+\delta}}$
for 
$\PermutationSpace{\dLoneText}$, $\PermutationSpace{\dtauText}$ and
$\PermutationSpace{\dinvLoneText}$.
While \Cref{thm:RD_equivalence} indicates a source code for 
$\PermutationSpace{\dLinfText}$ can be transformed into source codes for other spaces
under both average-case and worst-case distortions, we develop explicit compression
schemes for each permutation spaces as the transformation of permutation
representations incur additional computational complexity and hence may not be desirable.

\subsubsection{Permutation space $\PermutationSpace{\dLinfText}$}
Given distortion $D_n=\BigTheta{n^\delta}$, we 
apply the sorting subsequences scheme in \Cref{sec:sort_to_quantize}
and
choose $m = D_n+1$, which ensures the
maximal distortion is no more than $D_n$, and $k=\floor{n/m}$, which indicates
\begin{align*}
    k m &= \floor{n/m} m = n + \BigO{n^\delta}
    \\
    \log m &= \delta \log n + \SmallO{1}
    \\
    \Delta(k,m) &= km \log m + \SmallO{km \log m}
    \\
    &= \delta n \log n + \BigO{n}
    .
\end{align*}

\subsubsection{Permutation spaces $\PermutationSpace{\dLoneText}$ and $\PermutationSpace{\dtauText}$}
Given distortion $D_n = \BigTheta{n^{1 + \delta}}$, we 
apply the sorting subsequences scheme in \Cref{sec:sort_to_quantize}
and choose
\begin{align*}
    m &= (1/\alpha) \floor{D_n/n} \leq D_n / (n \alpha)
    \\
    k &= \floor{n/m},
\end{align*}
then
\begin{align*}
    km &= n - \abs{\BigO{n^\delta}}
    \\
    D &\leq \alpha km^2 \leq D_n 
    \\
    \Delta(k,m) &= \delta n \log n - n \log (\alpha e) + \SmallO{n} 
    ,
\end{align*}
where the constant $\alpha$ depends on the distortion measure and
whether we are considering worst-case or average-case distortion, 
as shown in \eqref{eq:DtauW}, \eqref{eq:DtauA}, \eqref{eq:DLoneW} and \eqref{eq:DLoneA}, and is
summarized in \Cref{tab:alpha_table}.

\begin{table}[!b]
\centering
\caption{Values of $\alpha$ for different compression scenarios.}
\begin{tabular}[h!]{|c|c|c|}
    \hline
    $\alpha$ & average-case & worst-case
    \\
    \hline
    $\PermutationSpace{\dLoneText}$ 
    & 1/3 & 1/2
    \\
    \hline
    $\PermutationSpace{\dtauText}$
    & 1/4 & 1/2
    \\
    \hline
\end{tabular}
\label{tab:alpha_table}
\end{table}

\subsubsection{Permutation space $\PermutationSpace{\dinvLoneText}$}
Given distortion $D_n = \BigTheta{n^{1 + \delta}}$, we apply the component-wise
scalar quantization scheme in \Cref{sec:scalar_quantization} and choose the
quantization error of the coordinate with range $[0:k-1]$ $D^{(k)}$ to be
\begin{align*}
    D^{(k)} = \frac{kD}{(n+1)^2}
    ,
\end{align*}
then
\begin{align*}
    m_k &= \ceil{k / \left( 2+D^{(k)} +1\right)} 
    = \ceil{\frac{k(n+1)^2}{2kD_n + (n+1)^2}}
    \\
    &\leq \ceil{\frac{(n+1)^2}{2D_n}},
\end{align*}
and the overall distortion and the codebook size satisfy
\begin{align*}
    D &= \sum_{k=2}^n = \frac{(n-1)(n+2)}{(n+1)^2} D_n \leq D_n,
    \\
    \log \cardinality{C_n} 
    &= \sum_{k=2}^n \log m_k
    \leq n \log \ceil{\frac{(n+2)^2}{2D_n}} 
    \\
    &= (1-\delta)n \log n + \BigO{n}
    .
\end{align*}

\subsection{Compression in the small distortion regime}
\label{sec:compress_small}
In this section we provide compression schemes in the small distortion regime for
$\PermutationSpace{\dtauText}$ and $\PermutationSpace{\dinvLoneText}$, where for any
$a > 0, 0 < \delta < 1$, $D_n = a n^{\delta}$. 

\subsubsection{Permutation space $\PermutationSpace{\dtauText}$}
When $a \geq 1$, let $m = \floor{2a}$ and $k = \floor{n^\delta/m}$, then
\begin{align*}
    \Delta(k,m) 
    &= k \log m!  
    \\
    &\geq (n^\delta/m - 1) \log m!
    = \frac{\log \floor{2a}!}{\floor{2a}} n^\delta + O(1)
    .
\end{align*}
And the worst-case distortion is upper bounded by
\begin{equation*}
    km^2/2 \leq \frac{n^\delta m}{2} \leq a n^\delta = D_n
    .
\end{equation*}

When $0 < a < 1$, let $m = 2$ and $k = \floor{D_n/2}$, then
\begin{equation*}
    \Delta(k,m) = k \log m! = \floor{\frac{D_n}{2}} \log 2
    = \frac{a \log 2}{2} n^\delta + O(1)
    .
\end{equation*}
And the worst-case distortion is no more than
\begin{olequation*}
    km^2/2 \leq D_n
    .
\end{olequation*}

\subsubsection{Permutation space $\PermutationSpace{\dinvLoneText}$}
When $a > 1$, let 
\begin{align*}
    m_k = \begin{cases}
        k & k \leq n - \floor{n^\delta}
        \\
        \ceil{ \olbfrac{k}{2a-1} } & k > n - \floor{n^\delta} 
    \end{cases}
    ,\quad k = 2, \ldots, n
\end{align*}
then the distortion $D^{(k)}$ for each coordinate $k$ satisfies 
\begin{align*}
    D^{(k)} \leq
    \begin{cases}
        a & k \leq \ceil{n^\delta} 
        \\
        0 & k > \ceil{n^\delta} 
    \end{cases}
    ,
    k=2, 3, \ldots, n,
\end{align*}
and hence overall distortion is $\sum_{k=2}^n D^{(k)} = (\floor{n^\delta}) a \leq D_n$.
In addition, the codebook size 
\begin{align*}
    \cardinality{\CodeW}
    &= \prod_{k=2}^n m_k
    \leq 
    \left( 
        \olbfrac{1}{2a-1}
    \right)^{\floor{n^\delta}}
    \prod_{k=2}^n k .
\end{align*}
Therefore, 
\begin{olequation*}
    \log \cardinality{\CodeW}
    \leq 
    \log n! - \floor{n^\delta} \log (2a - 1) + \BigO{\log n}
    .
\end{olequation*}

When $a \leq 1$, let
\begin{equation*}
    m_k = \begin{cases}
        \ceil{ \olfrac{k}{3} } & k < \ceil{D_n}
        \\
        k & k \geq \ceil{D_n}
    \end{cases}
    ,\quad k = 2, \ldots, n
\end{equation*}
and apply uniform quantization on the coordinate $k$ of the inversion vector
with $m_k$ points, 
Then the distortion $D^{(k)}$ for each coordinate $k$ satisfies 
\begin{align*}
    D^{(k)} \leq
    \begin{cases}
        1 & k < \ceil{D_n}
        \\
        0 & k \geq \ceil{D_n}
    \end{cases}
    ,
    k=2, 3, \ldots, n,
\end{align*}
and hence overall distortion is $\sum_{k=2}^n D^{(k)} = \ceil{D_n} - 1 \leq D_n$.
In addition, the codebook size 
\begin{align*}
    \cardinality{\CodeW}
    &= \prod_{k=2}^n m_k
    \leq
    \prod_{k=2}^{\ceil{D_n}-1} {\oltfrac{k+3}{3}} 
    \prod_{k=\ceil{D_n}}^{n} k
    \\
    &=
    \frac{1}{3^{\ceil{D_n}-1}} 
    \ceil{D_n} (\ceil{D_n}+1) (\ceil{D_n}+2)
    \prod_{k=5}^{n-1} k 
    .
\end{align*}
Therefore, 
\begin{olequation*}
    \log \cardinality{\CodeW}
    \leq \log n! - \ceil{a n^\delta} \log 3 + \BigO{\log n}.
\end{olequation*}

\subsection{Compression in the large distortion regime}
\label{sec:compress_large}
In this section we provide compression schemes in the large distortion regime for
$\PermutationSpace{\dtauText}$ and $\PermutationSpace{\dinvLoneText}$, where for any
$0 < b < 1/2$, $D_n = b n^{2}$. 

\subsubsection{Permutation space $\PermutationSpace{\dtauText}$}
Let $k = \ceil{1/(2b)}$ and $m = \floor{n/k}$, 
then 
\begin{align*}
    \Delta(k,m) 
    &= k \log m! 
    \geq 
    k \log (n/k -1)!
    \\
    &\geq 
    k [n/k \log(n/k) - n/k \log e+ \BigO{\log n}]
    \\
    &=
    n \log (n/e) - n \log \ceil{{1}/{(2b)}} + \BigO{\log n}
    .
\end{align*}
Hence 
\begin{olequation*}
    \hr(D_n) = \log n! - \Delta(k,m) \leq \log \ceil{{1}/{(2b)}} + \BigO{\log n}.
\end{olequation*}
And the worst-case distortion is upper bounded by
\begin{equation*}
    km^2/2 
    \leq 
    \olbfrac{n^2}{2k}
    \leq 
    \olbfrac{n^2}{1/b}
    = b n^2
    .
\end{equation*}

\subsubsection{Permutation space $\PermutationSpace{\dinvLoneText}$}
Let 
\begin{olequation*}
    m_k = \ceil{\olbfrac{k}{4b(k-1)+1}}, \quad k = 2, \ldots, n.
\end{olequation*}
The distortion $D^{(k)}$ for each coordinate $k$ satisfies 
\begin{align*}
    D^{(k)} = \ceil{\frac{1}{2} \left(\frac{k}{m} - 1\right)}
    \leq \ceil{2b(k-1)}
    ,
    k=2, 3, \ldots, n,
\end{align*}
and hence overall distortion
\begin{olequation*}
    \sum_{k=2}^n D^{(k)} 
    \leq 
    \sum_{k=2}^n 2b(k-1) + 1
    \leq 
    (b+1/n)n(n-1) 
\end{olequation*}.
In addition, the codebook size 
\begin{align*}
\cardinality{\CodeW}
    = \prod_{k=2}^n m_k
    \leq \prod_{k=2}^n \ceil{\frac{k-1}{4b(k-1)}}
    \leq \ceil{\frac{1}{4b}}^{n-1}
    .
\end{align*}
Therefore, 
\begin{olequation*}
    \log \cardinality{\CodeW} \leq n \log \ceil{\olbfrac{1}{4b}} + \BigO{1}
    .
\end{olequation*}

\section{Compression of permutation space with Mallows model}
\label{sec:mallows}
In this section we depart from the uniform distribution assumption and investigate the compression of a
permutation space with a non-uniform model---Mallows model~\cite{mallows_non-null_1957}, a 
model with a wide range of applications such as ranking, partial ranking, and even algorithm analysis 
(see \cite[Section 2e]{diaconis_analysis_2000} and the references therein). 
In the context of storing user ranking data, the Mallows model (or more generally, the mixture of Mallows model)
captures the phenomenon that user rankings are often similar to each other.  In the application of approximate
sorting, the Mallows model may be used to model our prior knowledge that permutations that are similar to the
reference permutation are more likely.

\begin{definition}[Mallows model]
\label{def:mallows}
We denote a \emph{Mallows model} with reference permutation (mode) $\pi$ and parameter $q$ as
$\Mallows{\pi}{q}$, 
where for each permutation $\sigma \in \PermutationSet$, 
\begin{align*}
    \Prob{\sigma; \Mallows{\pi}{q}} = \frac{q^{\dtau{\sigma}{\pi}}}{Z_{q,\pi}}
    ,
\end{align*}
where normalization $Z_{q,\pi} = \sum_{\sigma\in\PermutationSet} p^{\dtau{\sigma}{\pi}}$.
In particular, when the mode $\pi = \idperm$, 
$Z_q \defeq Z_{q,\idperm} = [n]_q!$~\cite[(2.9)]{diaconis_analysis_2000},
where $[n]_q!$ is the $q$-factorial
\begin{olequation*}
    [n]_q! = [n]_q [n-1]_q \ldots [1]_q
\end{olequation*}
and $[n]_q$ is the $q$-number 
\begin{equation*}
    [n]_q \defeq \begin{cases}
        \frac{1-q^n}{1-q} & q \neq 1
        \\
        n & q  = 1
    \end{cases}
    .
\end{equation*}
\end{definition}

As we shall see, the entropy of the permutation space with a Mallows model is in general $\BigTheta{n}$, implying
lower storage space requirement and \emph{potentially} lower query complexity for sorting. Since the Mallows model is
specified via the \ktdis, we use \ktdis{} as the distortion measure, and focus our attention on the average-case
distortion.

Noting the \ktdis{} is right-invariant~\cite{deza_metrics_1998}, for the purpose of compression,
we can assume the mode $\pi = \idperm$ without loss of generality, and denote the Mallows model by $\MallowsId{q} \defeq
\Mallows{\idperm}{q}$.

\subsection{Repeated insertion model}
The Mallows model can be generated through a process named \emph{repeated insertion model} (RIM),
which is introduced in~\cite{doignon_repeated_2004} and later applied
in~\cite{lu_learning_2011}.
\begin{definition}[Repeated insertion model]
\label{def:RIM}
Given a reference permutation $\pi \in \PermutationSet$ and a set of insertion probabilities
$\Set{p_{i,j}, 1 \leq i \leq n, 1 \leq j \leq i}$, RIM generates a new output $\sigma$ 
by repeated inserting $\pi(i)$ before the $j$-th element in $\sigma$ with probability $p_{i,j}$
(when $j=i$, we append $\pi(i)$ at the end of $\sigma$).
\end{definition}

\begin{remark}
    Note that the insertion probabilities at step $i$ is independent of the realizations of
    earlier insertions.
\end{remark}

The $i$-th step in the RIM process involves sampling from a multinomial distribution with 
parameter $p_{i,j}, 1 \leq j \leq i$.
If we denote the sampling outcome at the $i$-th step of the RIM process by $a_i, 1\leq i \leq n$,
then $a_i$ indicates the location of insertion.
By \Cref{def:RIM},
a vector $\ba = [a_1, a_2, \ldots, a_n]$ has an one-one correspondence to a
permutation, and we called this vector $\ba$ an \emph{insertion vector}. 

\begin{lemma}
\label{lemma:ins_inv_vector}
Given a RIM with reference permutation $\pi = \idperm$ and insertion vector
$\ins{\sigma}$,
then the corresponding permutation $\sigma$ satisfies
\begin{equation*}
\insx{\sigma}{i} = i - \einvx{\sigma}{i}
,
\end{equation*}
where $\einv{\sigma}$ is an \emph{extended inversion vector}, which simply is 
an inversion vector $\inv{\sigma}$ with 0 prepended.
\begin{align*}
    \einvx{\sigma}{i} = \begin{cases}
        0 & i = 1
        \\
        \invx{\sigma}{i-1} & 2 \leq i \leq n
    \end{cases}
\end{align*}
\end{lemma}

Therefore, 
\begin{align*}
    \dtau{\sigma}{\idperm} 
    &= \dinvLone{\sigma}{\idperm} 
    \\
    &= \sum_{i=1}^n (i - \insx{\sigma}{i})
    = \nchoosek{n+1}{2} - \sum_{i=1}^n \insx{\sigma}{i} 
    .
\end{align*}
\begin{example}
    For $n = 4$ and reference permutation $\idperm = [1,2,3,4]$, if $\ba =
    [1,1,1,1]$, then
    $\sigma = [4,3,2,1]$, which corresponds to $\einv{\sigma} = [0, 1, 2, 3]$.
\end{example}

\begin{theorem}[Mallows model via RIM~\cite{doignon_repeated_2004,lu_learning_2011}]
    \label{thm:Mallows_via_RIM}
    Given reference permutation $\pi$ and
    \begin{equation*}
        p_{i,j} = \frac{q^{i-j}}{1+q+\ldots+q^{i-1}}, 1 \leq j \leq i \leq n
        ,
    \end{equation*}
    RIM induces the same distribution as the Mallows model $\Mallows{\pi}{q}$.
\end{theorem}
This observation allows us to convert compressing the Mallows model to a standard
problem in source coding.
\begin{theorem}
    \label{thm:mallows_equivalence}
    Compressing a Mallows model is equivalent to compressing a vector source 
    $\bX = [X_1, X_2, \ldots, X_n]$, where $X_i$ is a geometric random
    variable truncated at $i-1, 1 \leq i \leq n$, \ie, 
    \begin{align*}
        \Prob{X_i = j} 
        &= \frac{q^{j}}{\sum_{j'=0}^{i-1} q^{j'}}
        \\
        &= \frac{q^{j}(1-q)}{1-q^i}
        ,
        0 \leq j \leq i-1
    \end{align*}
\end{theorem}
\begin{IEEEproof}
    This follows directly from 
    \Cref{lemma:ins_inv_vector}
    and
    \Cref{thm:Mallows_via_RIM}.
\end{IEEEproof}

\subsection{Lossless compression}
We consider the lossless compression of Mallows model. 
\begin{corollary}
    \begin{equation*}
    \Entropy{\cM(q)} = \Entropy{\cM(1/q)}
    \end{equation*}
\end{corollary}
\begin{IEEEproof}
    This follows directly from \Cref{thm:Mallows_via_RIM}.
\end{IEEEproof}

\begin{figure}[!bt]
    \begin{center}
        \addplot{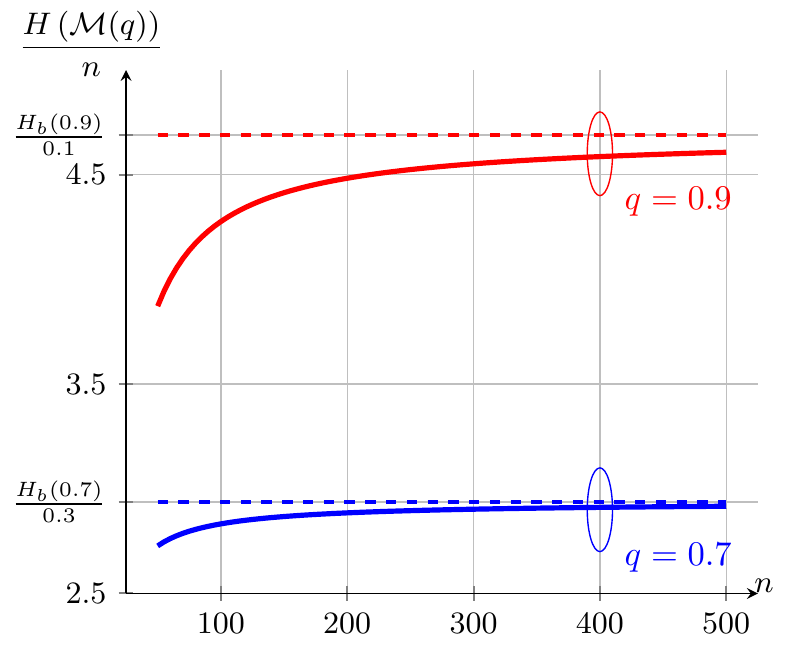}
    \end{center}
    \caption{Entropy of the Mallows model for $q=0.7$ and $q=0.9$, where the dashed lines
        are the coefficients of the linear terms, $\Hb{q}/(1-q)$.}
    \label{fig:mallows_entropy}
\end{figure}

\begin{lemma}[Entropy of Mallows model]
    \label{lemma:mallows_entropy}
\begin{align*}
    \Entropy{\cM(q)} &= \sum_{k=1}^n \Entropy{X_k}
    \\
    &= \begin{cases}
        \frac{\Hb{q}}{1-q} n + g(n,q)
        & q \neq 1
        \\
        \log n! & q = 1
    \end{cases}
    ,
\end{align*}
where $\Set{X_k}$ are truncated geometric random variables defined in \Cref{thm:mallows_equivalence},
$\Hb{\cdot}$ is the binary entropy function,
$g(n,q) = \BigTheta{1}$, and
$\lim_{q \rightarrow 0}g(n, q) = 0$.
\end{lemma}
The proof is presented in \Cref{sec:proof_mallows_entropy}. \Cref{fig:mallows_entropy}
shows plots of $\Entropy{\cM(q)}$ for different values of $n$ and $q$.

\begin{remark}
    Performing entropy-coding for each $X_i, 1\leq i \leq n$ is sub-optimal in general as
    the overhead is $O(1)$ for each $i$ and hence $O(n)$ for $\bX$, which is on the
    same order of the entropy $\Entropy{\cM(q)}$  when $q\neq 1$.
\end{remark}


\subsection{Lossy compression}
By \Cref{thm:mallows_equivalence}, the lossy compression of Mallows model is equivalent to the lossy compression
of the independent non-identical source $\bX$.  However, it is unclear whether an analytical solution of the
rate-distortion function for this source can be derived, and below we try to gain some insights via
characterizing the typical set of the Mallows model in \Cref{lemma:mallows_typical}, which implies that at rate 0,
the average-case distortion is $\BigTheta{n}$, while under the uniform distribution, \Cref{thm:RD_funcs}
indicates that it takes $n \log n + \SmallO{n \log n}$ bits to achieve average-case distortion of
$\BigTheta{n}$.

\begin{lemma}[Typical set of Mallows model]
    \label{lemma:mallows_typical}
    There exists $c_0(q)$, a constant that depends on $q$ only, such that
    for any $r_0 \geq c_0(q) n$, 
    \begin{align*}
        \lim_{\ntoinf} \Prob{
        d_{\tau}\left( \idperm, \sigma \right) \leq r_0 ; \Mallows{\idperm}{q} 
        } = 1 
        .
    \end{align*}
\end{lemma}
The proof is presented in \Cref{sec:proof_mallows_typical}.

\begin{remark}
As pointed out in~\cite{doignon_repeated_2004}, Mallows model is only one specific distributional model that is induced by RIM.
It is possible to generalize our analysis above to other distributional models that are also induced by RIM.
\end{remark}

\section{Concluding Remarks}
\label{sec:conclu}
In this paper, we first investigate the lossy compression of permutations under
both worst-case distortion and average-case distortions with uniform source distribution.
We consider \ktdis{}, Spearman's footrule, Chebyshev distance and
\invLoneName\ as distortion measures. 
Regarding the lossy storage of ranking, our results provide the fundamental trade-off between storage and
accuracy.  Regarding approximate sorting, our results indicate that, given a moderate distortion $D_n$ (see
\Cref{sec:rd_formulation} for definition), an approximate sorting algorithm must perform \emph{at least}
$\BigTheta{n \log n}$ pairwise comparisons, where constant implicitly in the $\Theta$ term is exactly the
rate-distortion function $\RateG(D_n)$. As mentioned, this performance is indeed achieved by the multiple
selection algorithm in~\cite{kaligosi_towards_2005}. This shows our information-theoretic lower bound for
approximate sorting is tight.

In practical ranking systems where prior knowledge on the ranking is available, non-uniform model may be more
appropriate. Our results on the Mallows model show that the entropy could be much lower ($\BigTheta{n}$) than
the uniform model ($\BigTheta{n \log n}$).  This greater compression ratio suggests that it would be worthwhile
to solve the challenge of designing entropy-achieving compression schemes with low computational complexity for
Mallows model. A deeper understanding on the rate-distortion trade-off of non-uniform models would be
beneficial to the many areas that involves permutation model with a non-uniform distribution, such as the
problem of learning to rank~\cite{lu_learning_2011} and algorithm analysis~\cite{diaconis_analysis_2000}.

\section*{Acknowledgment}
\addcontentsline{toc}{section}{Acknowledgment}
The authors are grateful to an anonymous reviewer whose comment prompted an important correction to an earlier version of this paper.

\begin{appendices}
\crefalias{section}{app}
\crefalias{subsection}{app}
\section{Geometry of permutation spaces}
\label{sec:geometry}
In this section we provide results on the geometry of the permutation
space that are useful in deriving rate-distortion bounds.

We first define $D$-balls centered at $\sigma \in \PermutationSet$ with radius $D$ under
distance $d(\cdot, \cdot)$ and their maximum sizes:
\begin{align}
    B_d(\sigma, D) 
    &\defeq \SetDef{\pi}{ d(\pi, \sigma) \leq D }
    ,
    \label{eq:D_ball_def}
    \\
    N_d(D) 
    &\defeq \max_{\sigma \in \cS_n} \cardinality{B_d(\sigma, D)}
    \label{eq:max_D_ball_size_def}
    .
\end{align}

Let $\ktBall[\sigma,D]$, $\LoneBall[\sigma,D]$ and $\invLoneBall[\sigma,D]$ be the balls that
correspond to the \ktdis{}, $\LoneText$ distance of the permutations, and $\LoneText$ distance of the inversion vectors,
and $\ktBallSize$, $\LoneBallSize$, and $\invLoneBallSize$ be their maximum sizes respectively.

Note that \eqref{eq:kt_bounds_l1inv} implies 
$
    \ktBall[\sigma,D]
    \subset
    \invLoneBall[\sigma,D]
$
and thus
$
    \ktBallSize
    \leq
    \invLoneBallSize
$.
Below we establish upper bounds for $\invLoneBallSize$ and $\ktBallSize$, which are
useful for establishing converse results later.

\begin{lemma}
    \label{lemma:ktBallSize_ub}
    For $0 \leq D \leq n$, 
    \begin{equation}
        \ktBallSize \leq \nchoosek{n+D-1}{D}.
        \label{eq:ktBallSize_ub}
    \end{equation}
\end{lemma}
\begin{IEEEproof}
    Let the number of permutations in $\PermutationSet$ with at most $k$
    inversions be $T_n(d) \defeq \sum_{k=0}^d K_n(k)$, where $K_n(k)$ is defined in 
    \eqref{eq:K_nk}. 
    Since $\PermutationSpace{d_\tau}$ is a regular metric space, 
    \begin{equation*}
        \ktBallSize = T_n(D),
    \end{equation*}
    which is noted in several references such as~\cite{knuth_art_1998}.
    An expression for $K_n(D)$ (and thus $T_n(D)$) for $D\le n$ appears in
    \cite{knuth_art_1998} (see \cite{barg_codes_2010} also). The following
    bound is weaker but sufficient in our context.

    By induction, or~\cite{shreevatsa_on-line_2013}, 
    $
        T_n(D) = K_{n+1}(D)
    $ when $D \leq n$.
    Then noting that for $k<n$, $K_n(k) = K_n(k-1) + K_{n-1}(k)$~\cite[Section 5.1.1]{knuth_art_1998} and for any $n \geq 2$,
    \begin{align*}
        K_n(0) &= 1,
        \quad
        K_n(1) = n-1, \quad
        K_n(2) = \nchoosek{n}{2} - 1
        ,
    \end{align*}
    by induction, we can show that when $1 \leq k < n$,
    \begin{equation}
        K_n(k) \leq \nchoosek{n+k-2}{k}.
        \label{eq:K_ub}
    \end{equation}
\end{IEEEproof}

The product structure of $\PermutationSpace{\dinvLoneText}$ leads to a simpler analysis of the upper bound of 
$\invLoneBallSize$.
\begin{lemma}
    \label{lemma:invball_ub}
    For $0 \leq D \leq n(n-1)/2$, 
    \begin{equation}
        \label{eq:invLoneBallSize_ub} 
        \invLoneBallSize 
        \leq
        2^{\min\Set{n,D}} \nchoosek{n+D}{D}
        .
    \end{equation}
\end{lemma}
\begin{IEEEproof}
    For any $\sigma \in \PermutationSet$, let $\bx = \inv{\sigma} \in \InvVectorSet$, then
    \begin{equation*}
        \cardinality{\invLoneBall}
        = \sum_{r=0}^D \cardinality{\SetDef{\by\in\InvVectorSet}{\dLone{\bx}{\by} = r}}
        .
    \end{equation*}
    Let $\bd \defeq \abs{\bx - \by}$, and $Q(n,r)$ be the number of integer solutions of the equation $z_1 + z_2 + \ldots + z_{n} = r$ with 
    $z_i \geq 0, 0 \leq i \leq n$, then 
    it is well known~\cite[Section 1.2]{stanley_enumerative_1997} that
    \begin{equation*}
        Q(n,r) = \nchoosek{n+r-1}{r}
        ,
    \end{equation*}
    and it is not hard to see that the number of such $\bd = [d_1, d_2, \ldots, d_{n-1}]$ that
    satisfies $\sum_{i=1}^{n-1}d_i = r$ is upper bounded by $Q(n-1,r)$.
    Given $\bx$ and $\bd$, at most $m \defeq \min\Set{D, n}$ elements in $\Set{y_i, 0 \leq i \leq n}$ correspond to $y_i = x_i \pm d_i$. Therefore, for any $\bx$,
    \begin{olequation*}
        \cardinality{\SetDef{\by\in\InvVectorSet}{\dLone{\bx}{\by} = r}} \leq 2^{m} Q(n,r)
    \end{olequation*}
    and hence
    \begin{equation*}
        \cardinality{B_{\LoneText}(\bx, D)} 
        \leq \sum_{r=0}^D 2^{m} Q(n,r)
        = 2^{m} \nchoosek{n+D}{D}. 
    \end{equation*}
\end{IEEEproof}

Below we upper bound $\log \ktBallSize$ and $\log \invLoneBallSize$ for small,
moderate and large $D$ regimes in
\Cref{lemma:logT_smallD,lemma:logT_moderateD,lemma:logT_largeD} respectively.
\begin{lemma}[Small distortion regime]
    \label{lemma:logT_smallD}
    When $D = a n^\delta, 0 < \delta \leq 1$ and $a>0$ is a constant, 
    \begin{align}
        \nn
        &\log \ktBallSize 
        \\
        &\leq
        \begin{cases}
            a ( 1 - \delta) n^\delta \log n + \BigO{ n^\delta },
            \quad 
            0 < \delta < 1
        \\
        n \left[ \log \frac{(1+a)^{1+a}}{a^a} \right] + \SmallO{n},
            \quad 
            \delta = 1
        \end{cases}
        ,
        \label{eq:logT_ub}
        \\
        &\log \invLoneBallSize
        \nn
        \\
        &\leq
        \begin{cases}
            a ( 1 - \delta) n^\delta \log n + \BigO{ n^\delta }
            ,\;\;
            0 < \delta < 1
        \\
            n \left[ 2 + \log \frac{(1+a)^{1+a}}{a^a} \right] + \SmallO{n}
            ,\;\;
            \delta = 1
        \end{cases}
        \hspace{-1ex}
        .
    \end{align}
\end{lemma}
\begin{IEEEproof}
    To upper bound $\ktBallSize$, when $0 < \delta < 1$, we apply Stirling's approximation to
    \eqref{eq:ktBallSize_ub} to have
    \begin{align*}
        &\log \nchoosek{n+D-1}{D} 
        \\
        &= 
        n \log \frac{n-1+D}{n-1}
        +
        D \log \frac{n-1+D}{D}
        + 
        \BigO{\log n}
        .
    \end{align*}
    Substituting $D = a n^\delta$, we obtain \eqref{eq:logT_ub}.
    When $\delta=1$, the result follows from (9) in~\cite[Section 4]{louchard_number_2003}.
    The upper bound on $\invLoneBallSize$ can be obtained similarly via \eqref{eq:invLoneBallSize_ub}.
\end{IEEEproof}

\begin{lemma}[Moderate distortion regime]
    \label{lemma:logT_moderateD}
    Given $D = \BigTheta{n^{1 + \delta}}$, $0 < \delta \leq 1$, then 
    \begin{equation}
        \label{eq:logT_moderateD}
        \log \ktBallSize 
        \leq
        \log \invLoneBallSize
        \leq
        \delta n \log n + \BigO{n}
        .
    \end{equation}
\end{lemma}
\begin{IEEEproof}
    Apply Stirling's approximation to \eqref{eq:invLoneBallSize_ub} and 
    substitute $D=\BigTheta{n^{1+\delta}}$.
\end{IEEEproof}
\begin{remark}
    It is possible to obtain tighter lower and upper bounds for $\log \ktBallSize$ 
    and $\log \invLoneBallSize$ based on results in \cite{barg_codes_2010}. 
\end{remark}

\begin{lemma}[Large distortion regime]
    \label{lemma:logT_largeD}
    Given $D = b n (n-1) \in \pints$, then
    \begin{equation}
        \label{eq:invLoneBallSize_large_ub}
        \log \ktBallSize \leq 
        \log \invLoneBallSize \leq 
            n \log (2ben) + \BigO{\log n}
            .
    \end{equation}
\end{lemma}
\begin{IEEEproof}
    Substitute $D=bn(n-1)$ into \eqref{eq:invLoneBallSize_ub}.
\end{IEEEproof}

\subsection{Proof of \eqref{eq:linf_lb_lone_prob}}
\label{sec:proof_linf_lb_lone_prob}
\begin{lemma}
    For any $\pi \in \PermutationSet$, let $\sigma$ be a permutation chosen uniformly from
    $\PermutationSet$, and
    $X_{\ell_1} \defeq \dLone{\pi}{\sigma}$, then
    \begin{align}
        \label{eq:mean_l1}
        \E{X_{\ell_1}} &= \frac{n^2-1}{3} 
        \quad
        \Var{X_{\ell_1}} = \frac{2n^3}{45} + \BigO{n^2}.
    \end{align}
    \label{lemma:l1_moments}
\end{lemma}
\begin{IEEEproof}
    \begin{align*}
        \E{X_{\ell_1}} &= \frac{1}{n} \sum_{i=1}^n \sum_{j=1}^n \abs{i-j}
        = \frac{2}{n} \sum_{i=1}^n \sum_{j=1}^i \abs{i-j}
        \\
        &= \frac{2}{n} \sum_{i=1}^n \sum_{j'=0}^{i-1} j'
        = \frac{1}{n} \sum_{i=1}^n (i^2 - i)
        \\
        &= \frac{1}{n} \left(\sum_{i=1}^n i^2 - \sum_{i=1}^n i\right)
        \\
        &= \frac{1}{n} \left( \frac{2n^3 +3n^2 + n}{6} - \frac{n^2+n}{2} \right)
        \\
        &= \frac{n^2-1}{3}.
    \end{align*}
    And $\Var{X_{\ell_1}}$ can be derived similarly~\cite[Table 1]{diaconis_spearmans_1977}. 
\end{IEEEproof}

\begin{IEEEproof}[Proof for \eqref{eq:linf_lb_lone_prob}]
    For any $c > 0$, 
    \begin{olequation*}
        c n \cdot \dLinf{\pi}{\sigma} 
        \leq c n(n-1) 
        ,
    \end{olequation*}
    and for any $c_1 < 1/3$, \Cref{lemma:l1_moments} and Chebyshev inequality indicate
    \begin{olequation*}
        \Prob{ \dLone{\pi}{\sigma} < c_1 n(n-1) } = O(1/n).
    \end{olequation*}
    Therefore, 
    \begin{align*}
        &\;{}\Prob{ \dLone{\pi}{\sigma} \geq c_1 n \cdot \dLinf{\pi}{\sigma} }
        \\
        \geq&\;{}
        \Prob{ \dLone{\pi}{\sigma} \geq c_1 n (n-1) }
        \\
        =&\;{} 1 - \Prob{ \dLone{\pi}{\sigma} < c_1 n (n-1) }
        \\
        =&\;{} 1 - \BigO{1/n}.
    \end{align*}
\end{IEEEproof}

\subsection{Proof of \Cref{thm:kt_lb_l1}}
\label{sec:proof_kt_bounds_l1}

\begin{lemma}
    For any two permutations $\pi, \sigma$ in $\PermutationSet$ such that 
    $\dinvLone{\pi}{\sigma} = 1$, 
    \begin{olequation*}
        \dtau{\pi}{\sigma} \leq n-1.
    \end{olequation*}
    \label{lemma:max_kt_at_inv1}
\end{lemma}
\begin{IEEEproof}
    Let 
    $\inv{\pi} = [a_2, a_3, \ldots, a_n]$ and 
    $\inv{\sigma} = [b_2, b_3, \ldots, b_n]$, then without loss of generality, we have
    for a certain $2 \leq k \leq n$, 
    \begin{equation*}
        a_i = \begin{cases}
            b_i & i \neq k
            \\
            b_i+1 & i = k
            .
        \end{cases}
    \end{equation*}
    Let $\pi'$ and $\sigma'$ be permutations in $\PermutationSet[n-1]$ with element $k$ removed
    from $\pi$ and $\sigma$ correspondingly, then 
    $\inv{\pi'} = \inv{\sigma'}$, and hence $\pi' = \sigma'$. Therefore, the \ktdis{} between
    $\sigma$ and $\pi$ is determined only by the location of element $k$ in $\sigma$ and $\pi$,
    which is at most $n-1$. 
\end{IEEEproof}

\begin{IEEEproof}[Proof of \Cref{thm:kt_lb_l1}]
    It is known that (see, e.g.,\cite[Lemma 4]{mazumdar_constructions_2013})
    \begin{equation*}
        d_{\ell_1}(\inv{\pi_1}, \inv{\pi_2}) \leq d_{\tau}(\pi_1, \pi_2).
    \end{equation*}
    Furthermore, 
    the proof of \cite[Lemma 4]{mazumdar_constructions_2013} indicates that for any two
    permutation $\pi_1$ and $\pi_2$ with $k = \dinvLone{\pi_1}{\pi_2}$, 
    let $\sigma_0 \defeq \pi_1$ and $\sigma_{k} \defeq \pi_2$, then 
    there exists a sequence of permutations 
    $\sigma_1, \sigma_2, \ldots, \sigma_{k-1}$ such that 
    $\dinvLone{\sigma_{i}}{\sigma_{i+1}} = 1, 0 \leq i \leq k-1$. 
    Then 
    \begin{align*}
        \dtau{\pi_1}{\pi_2} 
        &\leq
        \sum_{i=0}^{k-1} \dtau{\sigma_i}{\sigma_{i-1}}
        \\
        &\overset{(a)}{\leq}
        \sum_{i=0}^{k-1} (n-1) 
        = (n-1) \dinvLone{\pi_1}{\pi_2}
        ,
    \end{align*}
    where (a) is due to \Cref{lemma:max_kt_at_inv1}.
\end{IEEEproof}

\subsection{Proof of \eqref{eq:kt_lb_l1inv_prob}}
\label{sec:proof_kt_bounds_l1_prob}

To prove \eqref{eq:kt_lb_l1inv_prob}, we analyze the mean and variance of the \ktdis{} and
\invLoneName\ 
between a permutation in $\PermutationSet$ and a
randomly selected permutation, in \Cref{lemma:kt_moments} and \Cref{lemma:invLone_moments}
respectively. 

\begin{lemma}
    For any $\pi \in \PermutationSet$, let $\sigma$ be a permutation chosen uniformly from
    $\PermutationSet$, and
    $X_{\tau} \defeq \dtau{\pi}{\sigma}$, then
    \begin{align}
        \label{eq:E_kt_dis}
        \E{X_{\tau}} &= \frac{n(n-1)}{4},
        \\
        \Var{X_{\tau}} &= \frac{n(2n+5)(n-1)}{72} .
    \end{align}
    \label{lemma:kt_moments}
\end{lemma}
\begin{IEEEproof}
    Let $\sigma'$ be another permutation chosen independently and uniformly from
    $\PermutationSet$, then we have both $\pi\sigma^{-1}$  and $\sigma'\sigma^{-1}$ are
    uniformly distributed over $\PermutationSet$.

    Note that \ktdis{} is right-invariant~\cite{deza_metrics_1998}, 
    then
    $\dtau{\pi}{\sigma} = \dtau{\pi\sigma^{-1}}{\idperm}$
    and
    $\dtau{\sigma'}{\sigma} = \dtau{\sigma'\sigma^{-1}}{\idperm}$
    are identically distributed, 
    and hence the result follows~\cite[Table 1]{diaconis_spearmans_1977}
    and~\cite[Section 5.1.1]{knuth_art_1998}.
\end{IEEEproof}

\begin{lemma}
    For any $\pi \in \PermutationSet$, let $\sigma$ be a permutation chosen uniformly from
    $\PermutationSet$, and
    $X_{\invLoneText} \defeq \dinvLone{\pi}{\sigma}$, then
    \begin{align*}
        \E{X_{\invLoneText}} &> \frac{n(n-1)}{8},
        \\
        \Var{X_{\invLoneText}} &< \frac{(n+1)(n+2)(2n+3)}{6}
        .
    \end{align*}
    \label{lemma:invLone_moments}
\end{lemma}
\begin{IEEEproof}
    It is not hard to see that 
    when $\sigma$ is a permutation chosen uniformly from $\PermutationSet$, 
    $\inv{\sigma}(i)$ is uniformly distributed in $[0:i]$, $1\leq i \leq n-1$.
    Therefore, 
    \begin{olequation*}
        X_{\invLoneText} = \sum_{i=1}^{n-1} \abs{a_i - U_i},
    \end{olequation*}
    where $U_i \sim \PUniformSet{[0:i]}$ and $a_i \defeq \invx{\pi}{i}$. 
    Let $V_i = \abs{a_i - U_i}$, 
    $m_1 = \min\Set{i-a_i, a_i}$ and $m_2 = \max\Set{i-a_i, a_i}$, then
    \begin{align*}
        \Prob{V_i = d} = \begin{cases}
            1/(i+1) & d = 0
            \\
            2/(i+1) & 1 \leq d \leq m_1
            \\
            1/(i+1) & m_1 + 1 \leq d \leq m_2
            \\
            0 & \text{otherwise}
            .
        \end{cases}
    \end{align*}
    Hence,
    \begin{align*}
        \E{V_i}
        &= \sum_{d=1}^{m_1} d \frac{2}{i+1} + \sum_{d=m_1+1}^{m_2} d \frac{1}{i+1} 
        \\
        &= \frac{
                2(1+m_1)m_1 + (m_2 + m_1 + 1)(m_2 - m_1)
        }
        {2(i+1)} 
        \\
        &= \frac{1}{2(i+1)} (m_1^2 + m_2^2 + i)
        \\
        &\geq \frac{1}{2(i+1)} \left( \frac{(m_1 + m_2)^2}{2} + i \right)
        = \frac{i(i+2)}{4(i+1)}
        > \frac{i}{4}
        ,
        \\
        &\Var{V_i}
        \leq \E{V_i^2}
        \leq \frac{2}{i+1} \sum_{d=0}^{i} d^2 
        \leq (i+1)^2
        .
    \end{align*}
    Then, 
    \begin{align*}
        \E{X_{\invLoneText}}
        &= \sum_{i=1}^{n-1} \E{V_i}
        > \frac{n(n-1)}{8}
        ,
        \\
        \Var{X_{\invLoneText}}
        &= \sum_{i=1}^{n-1} \Var{V_i}
        < \frac{(n+1)(n+2)(2n+3)}{6}
        .
    \end{align*}
\end{IEEEproof}

With \Cref{lemma:kt_moments} and \Cref{lemma:invLone_moments}, now we show that the
event that a scaled version of the \ktdis{} is larger than the \invLoneName\ is
unlikely.

\begin{IEEEproof}[Proof for \eqref{eq:kt_lb_l1inv_prob}]
    Let $c_2 = 1/3$, let $t = \olfrac{n^2}{7}$, then noting
    \begin{align*}
        t 
        &= \E{c \cdot X_\tau} + \abs{\BigTheta{\sqrt{n}}} \Std{X_\tau}
        \\
        &= \E{X_{\invLoneText}} - \abs{\BigTheta{\sqrt{n}}} \Std{X_{\invLoneText}}
        ,
    \end{align*}
    by Chebyshev inequality, 
    \begin{align*}
        \Prob{ c \cdot X_\tau > X_{\invLoneText} }
        &\leq 
        \Prob{ c \cdot X_\tau > t}
        +
        \Prob{ X_{\invLoneText} < t}
        \\
        &\leq 
        \BigO{1/n} + 
        \BigO{1/n} 
        = \BigO{1/n}
        .
    \end{align*}
    The general case of $c_2 < 1/2$ can be proved similarly.
\end{IEEEproof}

\section{Proofs on rate-distortion functions}

\subsection{Proof of \Cref{thm:RD_equivalence}}
\label{sec:proof_RD_equivalence}
\begin{IEEEproof}
Statement 1 follows from \eqref{eq:inf_ub_l1}.

Statement 2 and 3 follow from \Cref{thm:l1_and_kt}. For statement 2, let the encoding mapping for 
the $(n, D_n)$ source code in $\PermutationSpace{\dLoneText}$ be $f_n$ and 
the encoding mapping in $\PermutationSpace{\dtauText}$ be $g_n$, then 
\begin{align*}
    g_n(\pi) = \left[ f_n(\invperm[\pi]) \right]^{-1}
\end{align*}
is a $(n, D_n)$ source code in $\PermutationSpace{\dtauText}$. The proof for Statement 3 is
similar.

Statement 4 follow directly from \eqref{eq:kt_bounds_l1inv}.

\end{IEEEproof}

\subsection{Proof of \Cref{thm:RD_funcs}}
\label{sec:proof_RD_funcs}
We prove \Cref{thm:RD_funcs} by achievability and converse. 

\subsubsection{Achievability}
The achievability for all permutation spaces of interest under both \emph{worst-case
distortion} and \emph{average-case distortion} are established via the explicit
code constructions in \Cref{sec:codes}.

\subsubsection{Converse}
For the converse, we show by contradiction that under average-case distortion, if the rate is less than
$1-\delta$, then the average distortion is larger than $D_n$. Therefore, $\RateA \geq 1 - \delta$, and hence
$\RateW \geq \RateA \geq 1 - \delta$.

When $\delta = 1$, $\RateA = \RateW = 0$. 
When $0 \leq \delta < 1$, for any $0 < \eps < 1 - \delta$ and any codebook $\CodeA$ with size such that
\begin{equation}
    \log \cardinality{\CodeA} = (1 - \delta - \eps) n \log n + \BigO{n}
    \label{eq:bad_codebook_size}
    ,
\end{equation}
from \eqref{eq:geq_rlships}, when $D_n = \BigTheta{n^{1+\delta}}$ or $D_n = \BigO{n}$, 
\begin{align*}
    \LoneBallSize[2D_n]
    \cardinality{\CodeA}
    \leq
    \ktBallSize[2D_n]
    \cardinality{\CodeA}
    &\leq
    \\
    \invLoneBallSize[2D_n]
    \cardinality{\CodeA}
    &\overset{(a)}{\leq}
    n!/2
    ;
\end{align*}
when $D_n = \BigTheta{n^{\delta}}$ or $D_n = \BigO{1}$, 
\begin{align*}
    \LinfBallSize[2D_n]
    \cardinality{\CodeA}
    \leq
    \LoneBallSize[2D_n n]
    \cardinality{\CodeA}
    &\leq
    n!/2
\end{align*}
when $n$ sufficiently large, where $(a)$ follows from \eqref{eq:logT_moderateD}.

Therefore, given $\CodeA$, there exists at least $n!/2$ permutations in
$\PermutationSet$ that has distortion larger than $2D_n$, and hence the average
distortion \wrt{} uniform distribution over $\PermutationSet$ is larger than $D_n$.

Therefore, for any codebook with size indicated in \eqref{eq:bad_codebook_size}, we have
average distortion larger than $D_n$. 
Therefore, any $(n, D_n)$ code must satisfy $\RateW \geq \RateA \geq 1 - \delta$.

\section{Proofs on Mallows Model}
\subsection{Proof of \Cref{lemma:mallows_entropy}}
\label{sec:proof_mallows_entropy}
\begin{IEEEproof}
When $q = 1$ the Mallows model reduces to the uniform distribution on the permutation space.
When $q \neq 1$, let $X^n = [X_1, X_2, \ldots, X_n]$ be the inversion vector,
and
denote a geometric random variable by $G$ and a
geometric random variable truncated at $k$ by $G_k$.
Define
\begin{align*}
    E_k = \begin{cases}
        0 & G \leq k
        \\
        1 & \text{o.w.}
    \end{cases}
    ,
\end{align*}
then $\Prob{E_k = 0} = Q_k = 1 - q^{k+1}$.
Note
\begin{align*}
    \Entropy{G_k, E} 
    &= \CondEntropy{G}{E_k} + \Entropy{E_k}
    \\
    &= \CondEntropy{E_k}{G} + \Entropy{G} 
    \\
    &= \Entropy{G} 
\end{align*}
and
\begin{align*}
    \CondEntropy{G}{E_k} 
    &= \CondEntropy{G}{E_k = 0}Q_k
    \\
    &\quad + \CondEntropy{G}{E_k = 1} (1-Q_k)
    \\
    &= \Entropy{G_k} Q_k + \Entropy{G} (1-Q_k)
    ,
\end{align*}
we have
\begin{align*}
    \Entropy{G_k}
    &= \Hb{q}/(1-q) - \Hb{Q_k}/Q_k
    .
\end{align*}
Then 
\begin{align*}
    \Entropy{\cM(q)} 
    &=
    \sum_{k=0}^{n-1} \Entropy{G_k}
    \\
    &= \frac{n \Hb{q}}{1-q} - \sum_{k=1}^{n} \frac{\Hb{q^k}}{1-q^k}
    .
\end{align*}
It can be shown via algebraic manipulations that 
\begin{align*}
    \sum_{k=1}^n \Hb{q^k} 
    &\leq \frac{2q - q^2}{(1-q)^2}
    = \BigTheta{1}
    ,
\end{align*}
therefore 
\begin{align*}
    \Entropy{\cM(q)} 
    &= \frac{n \Hb{q}}{1-q} - \BigTheta{1}
    .
\end{align*}

%
\end{IEEEproof}

\subsection{Proof of \Cref{lemma:mallows_typical}}
\label{sec:proof_mallows_typical}
We first show an upper bound $K_n(k)$ (\cf\ \eqref{eq:K_nk} for definition), the number of permutations with $k$
inversion in $\PermutationSet$.
\begin{lemma}[Bounds on $K_n(k)$]
    \label{lemma:bound_Kn}
    For $k = c n$, 
    \begin{align*}
        K_n(k) 
        \leq 
        \frac{1}{\sqrt{2\pi n c/(1+c)}}
        2^{n(1+c)\Hb{1/(1+c)}} 
        .
    \end{align*}
\end{lemma}
\begin{IEEEproof}
    By definition, $K_n(k)$ equals to the number of non-negative integer solutions 
    of the equation $z_1 + z_2 + \ldots + z_{n-1} = k$ with 
    $0 \leq z_i \geq i, 1 \leq i \leq n-1$. 
    Then similar to the derivations in the proof of \Cref{lemma:invball_ub}, 
    \begin{align*}
    K_n(k) < Q(n-1, k) = \nchoosek{n+k-2}{k}.
    \end{align*}
    Finally, applying the bound~\cite{cohen_covering_1997}
    \begin{equation*}
        \nchoosek{n}{pn} \leq \frac{2^{n\Hb{p}}}{\sqrt{2\pi n p (1-p)}}
    \end{equation*}
    completes the proof.
\end{IEEEproof}

\begin{IEEEproof}[Proof of \Cref{lemma:mallows_typical}]
    Note 
    \begin{align*}
        \dtau{\sigma}{\idperm} = \dinvLone{\sigma}{\mathbf{0}}
        .
    \end{align*}
    Therefore,
    \begin{align*}
        \sum_{\sigma \in \PermutationSet, \dtau{\sigma}{\idperm} \geq r_0} \Prob{\sigma}
        &= \frac{1}{Z_q} \sum_{r = r_0}^{\nchoosek{n}{2}} q^r K_n(r)
        .
    \end{align*}
    And \Cref{lemma:bound_Kn} indicates for any $r = cn$, 
    \begin{align*}
        q^{r} K_n(r)
        &\leq 
        \frac{ 2^{ n \left[ (1+c)\Hb{\frac{1}{1+c}} - c \log_2 \frac{1}{q}  \right] } }
        {\sqrt{2\pi n c /(1+c)}}
        .
    \end{align*}
    Define 
    \begin{align*}
        E(c, q) \defeq \left[ (1+c)\Hb{\frac{1}{1+c}} - c \log_2 \frac{1}{q}  \right] 
        ,
    \end{align*}
    then for any $\eps > 0$, there exits $c_0$ such that for any $c \geq c_0(q)$, $E(c, q) < -\eps$.  
    Therefore, let $r_0 \geq c_0 n$,
    \begin{align*}
        \sum_{\sigma \in \PermutationSet, \dtau{\sigma}{\idperm} \geq r_0} \Prob{\sigma}
        &\leq 
        \frac{1}{\sqrt{2\pi n c /(1+c)}}
        \frac{1}{Z_q} \sum_{r = r_0}^{\nchoosek{n}{2}} 2^{ - n \eps} 
        \\
        &\rightarrow 0 
    \end{align*}
    as $\ntoinf$.
\end{IEEEproof}

\end{appendices}


\begin{IEEEbiographynophoto}{Da~Wang}
received the B.A.Sc. degree with honors in electrical engineering from the University of Toronto, Toronto, ON,
Canada, and the S.M. and Ph.D. degrees in electrical engineering and computer science (EECS) from the
Massachusetts Institute of Technology (MIT), Cambridge, in 2008, 2010 and 2014, respectively.

Dr. Wang is a receipt for several awards or fellowships, including Jacobs Fellowship in
2008, Claude E. Shannon Research Assistantship in 2011-2012, and Wellington and Irene Loh Fund Fellowship in
2014. His research interests include information theory, distributed computing and statistical inference.
\end{IEEEbiographynophoto}

\begin{IEEEbiographynophoto}{Arya Mazumdar} (S'05-M'13)
is an assistant professor in University of Minnesota-Twin Cities (UMN) since January 2013. Before coming to UMN,
he was a postdoctoral scholar at the Massachusetts Institute of Technology (MIT). He received his Ph.D.  degree
from University of Maryland, College Park, in 2011. 

Arya is a recipient of 2014-15 NSF CAREER award and the 2010 IEEE ISIT Student Paper Award. He is also the
recipient of the Distinguished Dissertation Fellowship Award, 2011, at the University of Maryland.  He spent the
summers of 2008 and 2010 at the Hewlett-Packard Laboratories, Palo Alto, CA, and IBM Almaden Research Center,
San Jose, CA, respectively.  Arya's research interests include error-correcting codes, information theory and
their applications.
\end{IEEEbiographynophoto}

\begin{IEEEbiographynophoto}{Gregory~W.~Wornell} (S'83-M'91-SM'00-F'04)
received the B.A.Sc. degree in electrical engineering from the University of British
Columbia, Vancouver, BC, Canada, and the S.M. and Ph.D. degrees in electrical engineering and computer science
from the Massachusetts Institute of Technology, Cambridge, MA, in 1985, 1987, and 1991, respectively.

Since 1991, he has been on the faculty at MIT, where he is the Sumitomo Professor of Engineering in the
department of Electrical Engineering and Computer Science (EECS). He leads the Signals, Information, and
Algorithms Laboratory in the Research Laboratory of Electronics, and co-chairs the EECS department graduate
program. He has held visiting appointments at the former AT\&T Bell Laboratories, Murray Hill, NJ, the University
of California, Berkeley, CA, and Hewlett-Packard Laboratories, Palo Alto, CA.

His research interests and publications span the areas of information theory, digital communication, statistical
inference, and signal processing, and include algorithms and architectures for wireless networks, sensing and
imaging systems, multimedia applications, and aspects of computational biology and neuroscience.

Dr. Wornell has been involved in the Information Theory and Signal Processing Societies of the IEEE in a variety
of capacities, and maintains a number of close industrial relationships and activities. He has won a number of
awards for both his research and teaching.
\end{IEEEbiographynophoto}

\end{document}